\documentclass{article}

\usepackage{arxiv}

\usepackage[utf8]{inputenc} % allow utf-8 input
\usepackage[T1]{fontenc}    % use 8-bit T1 fonts
\usepackage{hyperref}       % hyperlinks
\usepackage{url}            % simple URL typesetting
\usepackage{booktabs}       % professional-quality tables
\usepackage{amsfonts}       % blackboard math symbols
\usepackage{nicefrac}       % compact symbols for 1/2, etc.
\usepackage{microtype}      % microtypography
\usepackage{lipsum}		% Can be removed after putting your text content
\usepackage{graphicx}
\usepackage[numbers]{natbib}
\usepackage{doi}

\usepackage{tikz}
\usepackage{pgfplots}
\usepackage{amssymb}
\usepackage{amsmath}

\usepackage{hyperref}
\usepackage{multirow} 
\usepackage{rotating}
\usepackage{amsmath}

\title{Packed Malware Detection Using Grayscale Binary-to-Image Representations} 

\author{%
\href{https://orcid.org/0000-0002-5840-6164}{\includegraphics[scale=0.06]{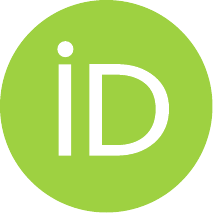}\hspace{1mm}Ehab Alkhateeb} \\
Canadian Institute for Cybersecurity (CIC)\\
University of New Brunswick\\
Fredericton, NB, Canada \\
\texttt{ehab.alkhateeb@unb.ca} \\
\And
\href{https://orcid.org/0000-0001-9189-6268}{\includegraphics[scale=0.06]{orcid.pdf}\hspace{1mm}Ali Ghorbani} \\
Canadian Institute for Cybersecurity (CIC)\\
University of New Brunswick\\
Fredericton, NB, Canada \\
\texttt{ghorbani@unb.ca} \\
\And
\href{https://orcid.org/0000-0002-1240-6433}{\includegraphics[scale=0.06]{orcid.pdf}\hspace{1mm}Arash Habibi Lashkari} \\
Behaviour-Centric Cybersecurity Center (BCCC)\\
York University\\
Toronto, ON, Canada \\
\texttt{ahabibil@yorku.ca} \\
}

\renewcommand{\shorttitle}{Packed Malware Detection Using Grayscale Binary-to-Image Representations}

\begin{document}
\maketitle

\begin{abstract}
Detecting packed executables is a critical step in malware analysis, as packing obscures the original code and complicates static inspection. This study evaluates both classical feature-based methods and deep learning approaches that transform binary executables into visual representations, specifically, grayscale byte plots, and employ convolutional neural networks (CNNs) for automated classification of packed and non-packed binaries. A diverse dataset of benign and malicious Portable Executable (PE) files, packed using various commercial and open-source packers, was curated to capture a broad spectrum of packing transformations and obfuscation techniques. Classical models using handcrafted Gabor jet features achieved intense discrimination at moderate computational cost. In contrast, CNNs based on VGG16 and DenseNet121 significantly outperformed them, achieving high detection performance with well-balanced precision, recall, and F1-scores. DenseNet121 demonstrated slightly higher precision and lower false positive rates, whereas VGG16 achieved marginally higher recall, indicating complementary strengths for practical deployment. Evaluation against unknown packers confirmed robust generalization, demonstrating that grayscale byte-plot representations combined with deep learning provide a useful and reliable approach for early detection of packed malware, enhancing malware analysis pipelines and supporting automated antivirus inspection.
\end{abstract}

% keywords can be removed

\label{subsec1}
\section{Introduction}
The rapid adoption of modern technologies, including smartphones, tablets, and wearable devices, has transformed how we access and share digital information, enabling seamless connectivity from virtually anywhere. Social networks, email, instant messaging, and IP telephony have further simplified information sharing, fostering unprecedented levels of connectivity. However, these technological advances have also facilitated the proliferation of computer malware, viruses, worms, Trojans, and other malicious software that pose serious threats to users, organizations, and critical infrastructure. Malware has increasingly become a strategic tool in cyber operations, often employed for espionage, sabotage, or other malicious purposes~\cite{jajodia2015cyber,Herrmann2019}.

Most executables encountered in malware analysis, both benign and malicious, are Portable Executable (PE) files targeting Windows systems. Malware authors frequently use packers, specialized tools that obfuscate, encrypt, or compress executables, to evade antivirus engines and hinder static analysis~\cite{lictua2018anti}. Many contemporary malware authors rely on sophisticated packers and obfuscation techniques to conceal the actual behavior of their malicious payloads~\cite{kaspersky2019ml}.

Packing transforms the original executable into a concealed form, often with a stub that manages runtime unpacking and may implement anti-analysis mechanisms. Detecting whether a file is packed is a crucial first step in analysis. Once a file is identified as packed, packer identification determines the specific type or family of the packer, enabling antivirus engines and analysts to apply targeted unpacking routines, recover the original code, and assess potential malicious behavior~\cite{alkhateeb2023survey}. Without these steps, malware analysis can be inefficient, error-prone, and computationally expensive.

The landscape of packers is vast and continually evolving, with many known, unknown, and custom variants employing sophisticated anti-analysis techniques. Efficient static analysis methods are therefore critical, offering a practical trade-off between accuracy and computational cost. Motivated by these challenges, this study proposes an image-based approach to malware analysis: binaries are converted into visual representations, such as Byte plots, and analyzed using machine learning and deep learning models, particularly convolutional neural networks (CNNs). This framework enables effective feature extraction from static binaries without requiring resource-intensive dynamic analysis, improving both accuracy and efficiency in detecting packed executables.

The widespread use of packing techniques complicates malware analysis, as traditional signature-based approaches struggle with unknown or sophisticated packers. Automated, reliable, and efficient methods are needed to detect packed files. This research addresses this gap by using image-based classification of PE files, providing a scalable and robust solution for real-world malware analysis.

To guide this investigation, we formulate the following research questions:

\begin{itemize}

\item {Can visual-based representations effectively distinguish packed from non-packed executables?} \\
Our results indicate that byte plot images capture distinct structural and statistical regularities, enabling reliable discrimination between packed and non-packed executables.

\item {How do classical feature-based models compare with deep learning models for packing detection?} \\
We observe that classical Gabor jet features offer competitive discrimination with low computational overhead; however, deep learning architectures such as VGG16 and DenseNet-121 consistently achieve superior accuracy and more balanced class-wise performance.

\item {Do models trained on known packers generalize to unknown or custom packing schemes?} \\
Our experiments demonstrate that models trained on diverse known packers generalize effectively, detecting a wide range of previously unseen packing transformations and exhibiting robust performance across heterogeneous packing techniques.

\item {What is the practical significance of early packing detection?} \\
We show that early identification of packed binaries facilitates efficient unpacking and dynamic analysis, minimizes manual reverse-engineering effort, and improves the throughput of antivirus inspection pipelines.
\end{itemize}

The contributions of this paper are summarized as follows:

\begin{itemize}
\item Construction of a comprehensive dataset comprising benign and malicious executables, including packed and non-packed samples generated with a variety of packers, as well as an adversarial dataset crafted to evaluate the robustness of the models.
\item Design and evaluation of classical machine learning models (e.g. Random Forests) and deep learning models (e.g. VGG16) to compare their effectiveness in distinguishing packed and non-packed binaries.
\item Evaluation of the robustness and generalization of the best-performing models, with an emphasis on their performance against unknown packers.
\end{itemize}

The remainder of this paper is organized as follows. Section \ref{sec:background} provides background on packers and malware packing techniques. Section \ref{sec:img} describes the binary-to-image conversion process. Section \ref{sec:gabor} details Gabor jet feature extraction for image-based classification of packed and non-packed executables. Section \ref{sec:dlm} presents the deep learning models used in this study, namely VGG16 and DenseNet121. Section \ref{sec:dataset} introduces the dataset and the motivation behind its construction. Section \ref{sec:experimental-setup} outlines the experimental setup and presents the results. Section \ref{sec:discussion} discusses the findings, implications, and limitations. Finally, Section \ref{sec:conc} concludes the study and highlights directions for future research.

\section{Background}\label{sec:background}
\subsection{Packers}
Packers are software tools designed to transform executable files. Their legitimate uses include protecting intellectual property from reverse engineering, reducing disk footprint, and lowering transmission time. However, packers are also commonly used by attackers for code obfuscation. When applied to malware, packing wraps the original program in an additional layer of code to hide its actual contents. Attackers may develop custom (malicious) packers or reuse legitimate packing tools for this purpose.

\subsection{How packing works}
\begin{figure*}[h!tbp]
\center
{\includegraphics[scale=0.56]{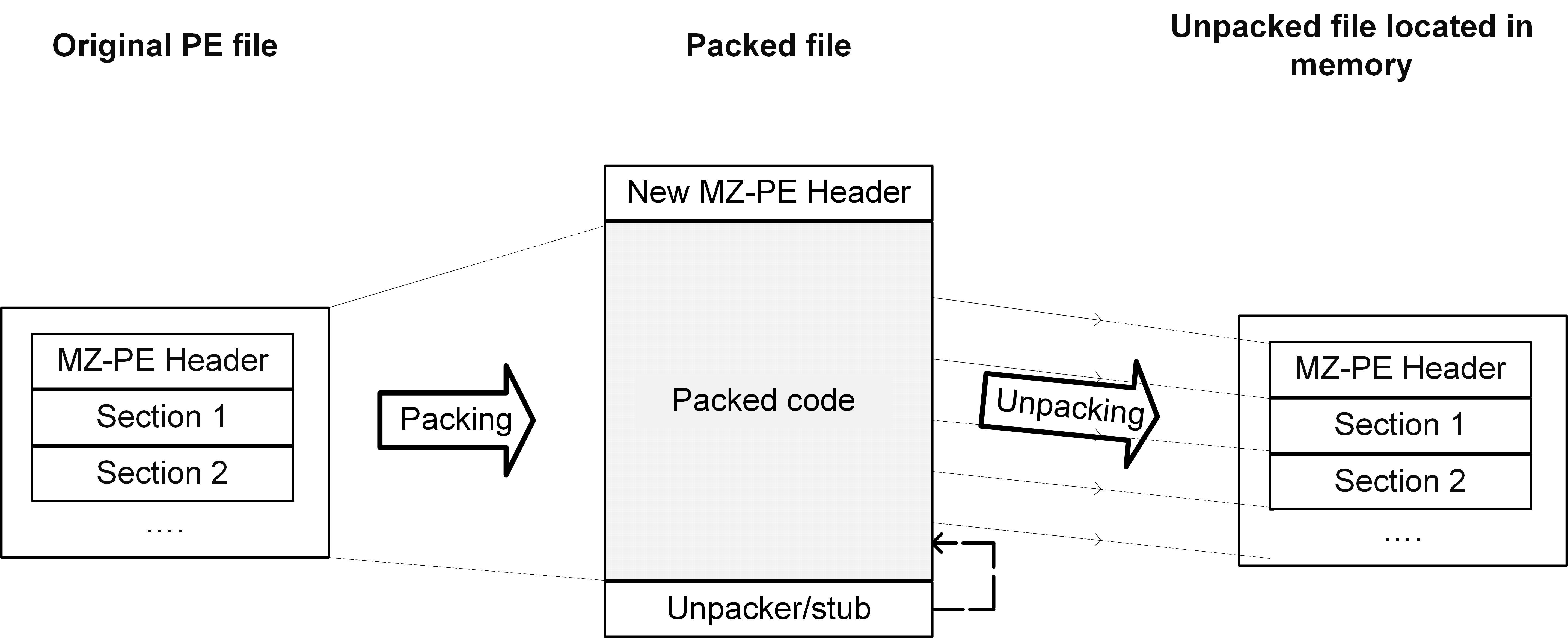}}
\caption{Packing ~\cite{alkhateeb2024identifying}.}
\label{fig:packing}
\end{figure*}

Packing is the process of using packers to obfuscate and, often, encrypt executables, malware included, thereby evading detection by security products and impeding manual analysis. A packer produces a transformed file that, at runtime, reconstructs the original code in memory. Figure~\ref{fig:packing} shows the transformation performed by a packer and the subsequent in-memory unpacking and code reconstruction.

A packed executable typically contains a small piece of code called a stub. The stub acts as the entry point when the packed file is executed and is responsible for driving the unpacking and decryption of the compressed or obfuscated payload. Its main roles include:

\begin{itemize}
\item {Initialization:} The stub is loaded into memory when the packed executable starts.

\item {Decryption and extraction:} The stub carries the logic needed to decrypt and extract the original payload, reversing the encryption and obfuscation applied by the packer.

\item {Execution:} After the payload is recovered, the stub transfers control to the reconstructed code so it can run.

\item {Anti-analysis measures:} Some stubs embed countermeasures that complicate static and dynamic analysis, making it harder for researchers and automated tools to inspect the hidden content.
\end{itemize}

The stub is therefore central to the packed executable’s operation: it enables the hidden payload to be unveiled and executed while preserving the obfuscation and evasion properties that make packing attractive to malware authors.
\begin{figure}[t]%% placement 
\centering%% For centre alignment of image.
\includegraphics[scale=0.11]{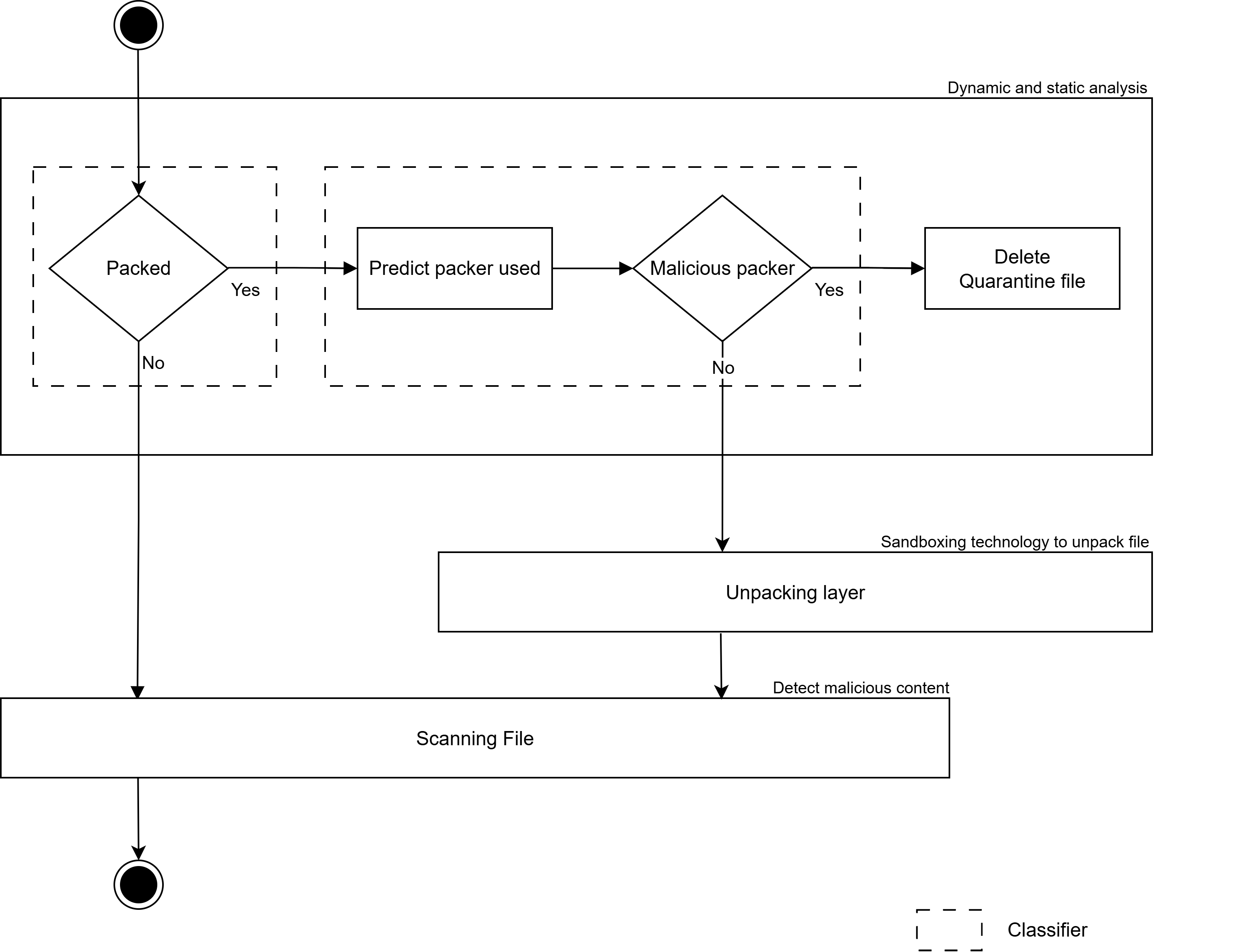}
\caption{AV engine workflow for packed files, emphasizing the binary packed/non-packed classification stage.}\label{figPNP}
\end{figure}
\subsection{Importance of determining packed vs. non-packed files}
Before attempting to classify the specific packer family, it is essential first to determine whether a given executable is packed~\cite{alkhateeb2024unmasking}. This distinction is critical because applying packer classification techniques to an unpacked file can lead to unnecessary computation, misclassification, or noise in the analysis. Figure~\ref{figPNP} illustrates the workflow of an AV engine when handling packers. The first stage, the focus of this paper, involves binary classification of a scanned sample as either packed or non-packed.

Packed files usually exhibit characteristics such as abnormal section sizes, high entropy values, and irregular import tables, features that differ significantly from those of benign or non-packed executables. Detecting these traits early enables analysts and automated systems to determine whether unpacking is required before deeper inspection~\cite{alkhateeb2023survey}.

The benefits of this preliminary step include:

\begin{itemize}
\item {Efficiency:} Avoids running packer-family identification and unpacking routines on files that are already unpacked.

\item {Accuracy:} Reduces false positives by ensuring that classification methods are applied only to actual packed samples.

\item {Resource Allocation:} Directs analysis resources (time, memory, and processing power) toward suspicious samples that warrant further unpacking and investigation.

\item {Workflow Optimization:} Establishes a clear decision point, whether to proceed with family identification, unpacking, or directly analyze the executable code.
\end{itemize}

In summary, determining whether a file is packed is a crucial preprocessing step that ensures subsequent packer-family classification is applied meaningfully and efficiently, improving the accuracy and reliability of the overall analysis pipeline.

\subsection{Machine learning models}
This study investigates eight machine learning models, spanning traditional algorithms, deep learning architectures, and pre-trained convolutional neural network frameworks. The following subsections provide a concise overview of each model’s key characteristics and application considerations.

\subsubsection{K-nearest neighbors}
The K-Nearest Neighbors (KNN) algorithm is a non-parametric, instance-based learning method that classifies a sample based on the majority class among its nearest neighbors in the feature space~\cite{cover1967nearest}. KNN does not make prior assumptions about the data distribution, making it suitable for a wide range of classification tasks. The choice of $k$ and the distance metric significantly impacts performance. Despite its simplicity, KNN can be computationally expensive for large datasets.

%\subsection{Naïve Bayes (NB)}
%Naïve Bayes (NB) is a probabilistic classifier based on Bayes' theorem, assuming independence among predictors~\cite{mccallum1998comparison}. It calculates the posterior probability of each class given the input features and assigns the label with the highest probability. NB performs well with high-dimensional data and is widely used for text classification and spam detection. However, its independence assumption may limit performance when features are correlated.

\subsubsection{Logistic regression}
Logistic Regression is a statistical classification method that models the probability of a binary outcome using a logistic (sigmoid) function. It estimates the relationship between input features and the likelihood of a class label by fitting a linear decision boundary in a transformed probability space \cite{hosmer2013applied}. Logistic Regression is computationally efficient, interpretable, and effective for linearly separable problems. However, its performance may degrade when complex nonlinear relationships are present in the data, unless feature engineering or kernel-based extensions are employed.

\subsubsection{Random forest}
Random Forest is an ensemble learning method that builds multiple decision trees and aggregates their predictions to improve classification accuracy~\cite{simonyan2014very}. It reduces variance compared to individual trees and is robust against overfitting. Each tree is trained on a randomly sampled subset of the data and features, thereby enhancing model diversity. Random Forests are effective for both classification and regression tasks in various domains.

\subsubsection{Support vector machine}
Support Vector Machine (SVM) is a supervised learning algorithm introduced by Cortes and Vapnik~\cite{cortes1995support}. It aims to find the optimal hyperplane that maximizes the margin between different classes in the feature space. SVM can efficiently handle high-dimensional data and employs kernel functions to model non-linear decision boundaries. Although computationally intensive for large datasets, SVMs often achieve high accuracy and generalization performance in classification tasks.

\subsubsection{The multilayer perceptron}
The Multilayer Perceptron (MLP) is a supervised feedforward neural network architecture composed of fully connected layers that learn nonlinear mappings between inputs and outputs. An MLP consists of an input layer, one or more hidden layers, and an output layer, with each neuron applying an activation function to enable the modeling of complex decision boundaries. As a general supervised learning framework, the MLP is trained by minimizing a task-specific loss function via gradient-based optimization, thereby enabling the network to refine its internal representations iteratively. Theoretical work has shown that, with sufficient capacity and appropriate activation functions, multilayer feedforward networks act as universal function approximators capable of representing arbitrarily complex relationships \cite{hornik1989multilayer}. In practice, however, MLP performance is often sensitive to initialization, feature scaling, and hyperparameter choices, which can result in training instability and noticeable variability across runs, particularly when applied to high-dimensional or structured data.

\subsubsection{Extreme gradient boosting}
Extreme Gradient Boosting (XGBoost) is a tree-based ensemble learning algorithm that leverages boosting to achieve high predictive accuracy and computational efficiency. XGBoost distinguishes itself from conventional boosting methods, such as AdaBoost, by incorporating advanced features including optimized split-finding algorithms, efficient caching, and parallel processing, which collectively enhance stability and performance. More broadly, boosting is a general learning framework in which multiple weak classifiers are iteratively combined to construct a stronger model; under ideal conditions, this approach can produce a classifier of virtually arbitrary predictive power~\cite{chen2016xgboost}.

\subsubsection{Transfer learning models}
Transfer learning leverages knowledge gained from large pre-trained convolutional neural networks, such as VGG16~\cite{simonyan2014very}, ResNet~\cite{he2016deep}, Inception~\cite{szegedy2016rethinking}, and DenseNet-121~\cite{huang2017densely}, to solve new tasks with limited training data. 
By reusing the early convolutional layers, which extract low-level patterns such as edges, textures, and frequency gradients, transfer learning significantly reduces computational cost and accelerates convergence compared with training from scratch.

This approach has become increasingly valuable for malware detection~\cite{brosolo2025through,kumar2022dtmic}, particularly in vision-based methods, where executable files are converted into grayscale or RGB images. Pre-trained models provide strong feature extraction capabilities that help capture structural differences between benign and malicious samples, even when only a modest number of labeled binaries are available. Such capability is especially critical in real-world cybersecurity settings, where malware evolves rapidly and collecting large, fully annotated datasets is often challenging. Transfer learning also offers robustness against obfuscation techniques such as packing. Packed malware usually exhibits distinctive entropy patterns, compressed regions, and abrupt texture transitions when rendered as images. Pre-trained CNNs can generalize these subtle visual cues more effectively than shallow or traditional machine-learning models, enabling detection of packed malware even when the underlying code is concealed.  Fine-tuning the network's deeper layers enables adaptation to domain-specific visual characteristics of executable files, improving classification accuracy and reducing overfitting despite dataset imbalance or noise.

\subsection{Related works}\label{RW}
Recent research on packer detection has explored diverse strategies for distinguishing packed executables from benign or unpacked applications. These approaches generally fall into two categories: \textit{dynamic analysis}, which observes runtime behavior in a controlled environment, and \textit{static analysis}, which examines file attributes without execution.

Dynamic analysis executes binaries in sandboxed or emulated environments to monitor behavior. Ugarte \textit{et al.}~\cite{ugarte2015sok} provided a comprehensive overview of generic unpackers, noting their fragility due to dependence on specific packer families. Hai \textit{et al.}~\cite{hai2017packer} combined metadata signatures with Control Flow Graph (CFG) analysis via concolic testing to detect obfuscation, though the method was computationally intensive. Alkhateeb \textit{et al.}~\cite{alkhateeb2019dynamic} and Menéndez \textit{et al.}~\cite{menendez2019mimicking} used API call analysis and entropy-based behavioral imitation, respectively, achieving high accuracy but limited scalability. Leal \textit{et al.}~\cite{leal2025low} leveraged Hardware Performance Counters during unpacking to classify low-entropy packers with machine learning. Entropy-based dynamic detection~\cite{bat2017entropy,bat2017packer2} and memory analysis frameworks such as Mal-Flux~\cite{lim2019mal} offered precise insights but at high runtime costs. Overall, dynamic approaches provide deep behavioral visibility but are resource-intensive, constraining their use in large-scale malware triage.

Static analysis extracts structural or statistical features from binaries without execution. Early works applied semi-supervised learning~\cite{ugarte2011semi} and PE header– or byte-level features~\cite{santos2011collective,naval2012escape} to detect packing artifacts. XOR- and entropy-based heuristics~\cite{laxmi2011peal,naval2015efficient} achieved high detection rates, while lightweight PE header–based metrics~\cite{jin2015packer,choi2008pe} were limited to known packers. Hybrid graph-based models~\cite{liu20212} and control-flow–based approaches~\cite{saleh2017control,li2019consistently} improved accuracy but encountered scalability challenges.

Visual analysis has emerged as an effective static paradigm. Byte-plot and Markov-plot methods~\cite{kancherla2016packer} combined texture feature extraction with SVM classification, while later works~\cite{jung2020packer,dam2022packer,biondi2019effective,bergenholtz2020detection,noureddine2021se,mimura2022applying} integrated statistical, rule-based, and neural approaches for multi-class detection. Deep learning approaches further enhanced feature representation, using CNNs on byte-frequency distributions~\cite{jung2020packer} and RNNs on instruction mnemonics~\cite{bergenholtz2020detection}. Self-evolving frameworks~\cite{noureddine2021se} incorporated clustering and online adaptation, though sequence-distance metrics introduced fragility across diverse packer families.

Despite prior work on static feature-based packer detection, many approaches rely on handcrafted attributes or shallow statistical patterns, which can limit generalization to unknown or novel packers. In this study, we address this limitation by converting binary images into Byte plot images and employing convolutional neural networks (CNNs) to extract discriminative spatial and structural patterns automatically. Gabor Jet features provide texture-based descriptors that capture local variations introduced by packing. Although their standalone classification performance is lower than that of CNN-based representations, they offer valuable insights into the local structural patterns of packed binaries. Overall, the results demonstrate that CNN-based Byte plot analysis provides a robust and scalable solution for detecting packed executables in automated malware analysis pipelines.

\section{Binary-to-image conversion}\label{sec:img}

To prepare binary executables for machine learning analysis, each file is first read in binary mode and converted into an array of 8-bit unsigned integers, with values ranging from 0 to 255. Each integer corresponds to a pixel intensity, with 0 representing black and 255 representing white. The bytes are arranged sequentially from left to right and top to bottom to construct a two-dimensional grayscale image, commonly referred to as a \emph{Byte plot}.

The image width can be set as a fixed parameter or determined adaptively based on the file size, while zero-padding is applied when the total number of bytes is not an exact multiple of the width, ensuring a complete rectangular grid. The height is computed automatically from the total number of bytes and the chosen width. This systematic mapping preserves both local and global structural patterns within the binary, allowing subtle variations introduced by packing transformations to become visually discernible.  

These generated grayscale images serve as input to convolutional neural networks (CNNs) and other supervised learning models, including approaches that leverage Gabor jet descriptors, thereby enabling automated extraction of both spatial and textural features. By transforming binaries into an image domain, the approach effectively bridges the gap between traditional static analysis and modern deep learning techniques, capturing patterns that are often difficult to detect using conventional feature engineering.  

\begin{figure}[t]%% placement 
\centering%% For centre alignment of image.
\includegraphics[scale=0.21]{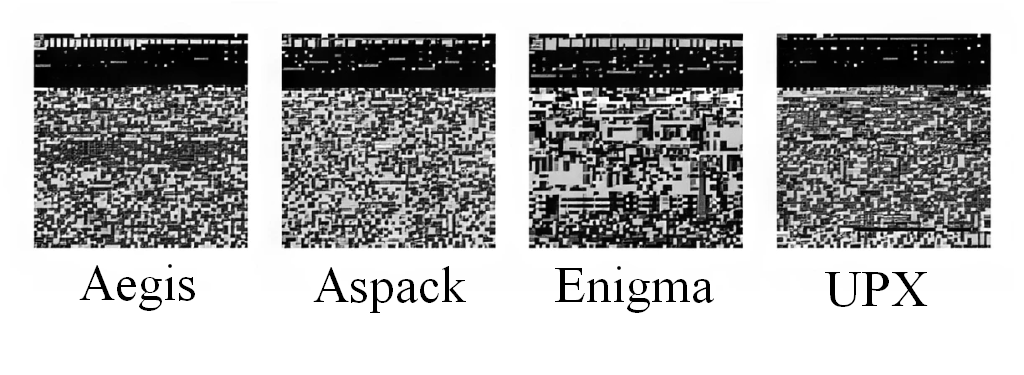}
\caption{Grayscale images of packed PE files generated using different packer families.}
\label{figcomb}
\end{figure}

Figure~\ref{figcomb} illustrates sample Byte plots generated from four different packed PE files, highlighting how various packer families produce distinctive visual patterns that image-based models can learn. This conversion step is central to the proposed pipeline, as it enables subsequent models to leverage powerful visual feature-extraction methods, thereby improving the accuracy and robustness of packed malware detection.  

Furthermore, the binary-to-image representation supports scalability and generalization across different file sizes and formats, providing a uniform input structure for deep learning pipelines. It also enables the integration of additional image-based preprocessing techniques, such as histogram equalization or noise reduction, to further enhance feature discriminability and classification performance.

\section{Gabor jet features}\label{sec:gabor}
Gabor filters are widely used in image processing for texture and edge analysis because they capture spatial-frequency and orientation information analogous to those of the human visual system. In malware analysis, such filters have been employed to characterize packed and unpacked executables using image-based representations~\cite{kancherla2016packer,alkhateeb2024identifying}. Mathematically, a two-dimensional Gabor filter is expressed as a sinusoidal plane wave modulated by a Gaussian envelope:

\begin{equation}
g(x, y) = \exp\left(-\frac{x'^2 + \gamma^2 y'^2}{2\sigma^2}\right) 
          \cos\left(2\pi \frac{x'}{\lambda} + \psi\right),
\end{equation}

where $x' = x \cos \theta + y \sin \theta$ and $y' = -x \sin \theta + y \cos \theta$. 
The parameters $\lambda$, $\theta$, $\psi$, $\sigma$, and $\gamma$ represent the wavelength, orientation, phase offset, standard deviation of the Gaussian envelope, and spatial aspect ratio, respectively. These parameters jointly determine the scale, direction, and selectivity of the filter response.

A \emph{Gabor jet} denotes the collective responses of multiple Gabor filters applied to a localized image region. Each jet encodes local texture and orientation information, enabling robust characterization of structural patterns in images derived from malware binaries. This representation is beneficial for discriminating between packed and unpacked executables, which exhibit distinct texture distributions when visualized as byte plots.
The feature extraction pipeline begins with image preprocessing, where each malware image is converted to grayscale and resized to $64 \times 64$ pixels to ensure uniform input dimensions. A bank of Gabor filters is then applied at multiple scales and orientations. In this study, filters were configured with three frequencies ($f \in \{0.1, 0.2, 0.3\}$) and four orientations ($\theta \in \{0, \pi/4, \pi/2, 3\pi/4\}$). The kernel size was set to $9\times9$, with Gaussian width $\sigma = 3.0$ and aspect ratio $\gamma = 0.5$.

Each filter response is obtained through a two-dimensional convolution between the image $I$ and the Gabor kernel $g$:

\begin{equation}
\text{Feature} = \left[\text{mean}(I * g), \text{var}(I * g)\right],
\end{equation}

where $*$ denotes convolution. The mean and variance of each filtered image quantify the strength and dispersion of texture energy at specific orientations and scales. The complete feature vector is formed by concatenating these values across all filter responses.

Given $n_f$ frequencies and $n_\theta$ orientations, each producing two statistical measures (mean and variance), the resulting feature dimensionality is defined as:

\begin{equation}
\text{Total features} = n_f \times n_\theta \times 2.
\end{equation}

For the chosen parameters ($n_f = 3$, $n_\theta = 4$), each image is represented by a 24-dimensional feature vector capturing multi-scale and multi-orientation texture characteristics.

Extracted features are stored in both compressed NumPy (\texttt{.npz}) format, containing matrices \texttt{X} (features) and \texttt{y} (labels), and in a tabular CSV format for ease of inspection and interoperability. These feature vectors can be directly used with classical machine learning models or fused with CNN-learned features to improve overall classification robustness.

\begin{figure}[h!]
\centering
\begin{tikzpicture}[auto, block/.style={rectangle, draw, text width=6cm, minimum height=1.5cm, align=center}]
\footnotesize
    % Nodes stacked vertically at x=0, with y-step of -1.75 to reduce space
    \node[block] (load) at (0,0) {Load images from benign and malicious folders};
    \node[block] (grayscale) at (0,-1.75) {Convert images to grayscale and resize to $64\times64$};
    \node[block] (gabor) at (0,-3.5) {Apply Gabor filter bank for multiple frequencies and orientations};
    \node[block] (extract) at (0,-5.25) {Extract mean and variance of each filter response};
    \node[block] (concat) at (0,-7) {Concatenate all features into a single vector per image};
    \node[block] (repeat) at (0,-8.75) {Repeat for all images and store feature matrix and labels};
    \node[block] (save) at (0,-10.5) {Save features in both \texttt{.npz} and CSV formats};

    % Draw straight vertical arrows
    \foreach \i/\j in {load/grayscale, grayscale/gabor, gabor/extract, extract/concat, concat/repeat, repeat/save}
        \draw[->, thick] (\i.south) -- (\j.north);

\end{tikzpicture}
\caption{Gabor feature extraction workflow.}
\label{fig:gabor_flow_reduced_spacing}
\end{figure}
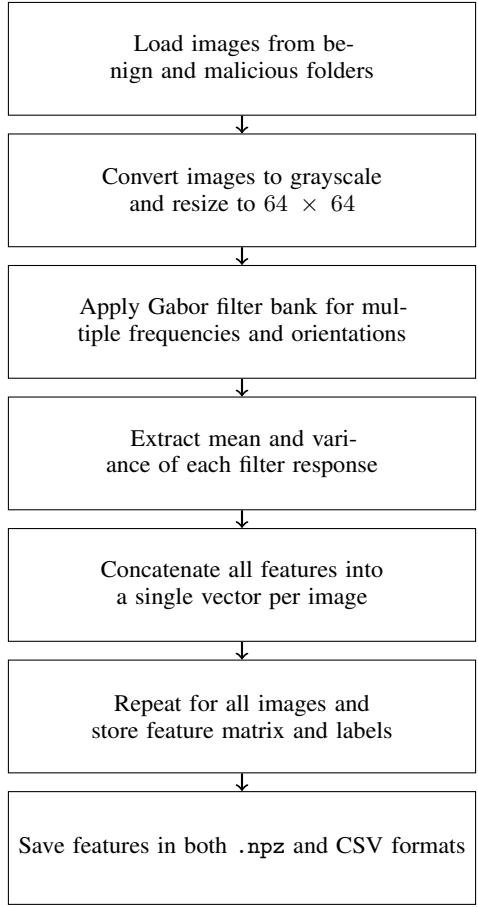

The Gabor Jet representation offers several advantages for malware image classification. It effectively captures fine-grained local texture and structural information while maintaining low dimensionality, reducing the risk of overfitting on limited datasets. Furthermore, the representation is computationally efficient and compatible with a wide range of supervised learning algorithms, including Random Forests, Support Vector Machines, and Logistic Regression. In contrast to deep CNN models, Gabor Jets require minimal training data and processing resources yet still provide strong discrimination between packed and unpacked executables.

Figure~\ref{fig:gabor_flow_reduced_spacing} summarizes the key stages of the Gabor-based feature extraction process used in this study.

\section{Deep learning models}\label{sec:dlm}
\begin{figure}[t]%% placement 
\centering%% For centre alignment of image.
\includegraphics[scale=0.09]{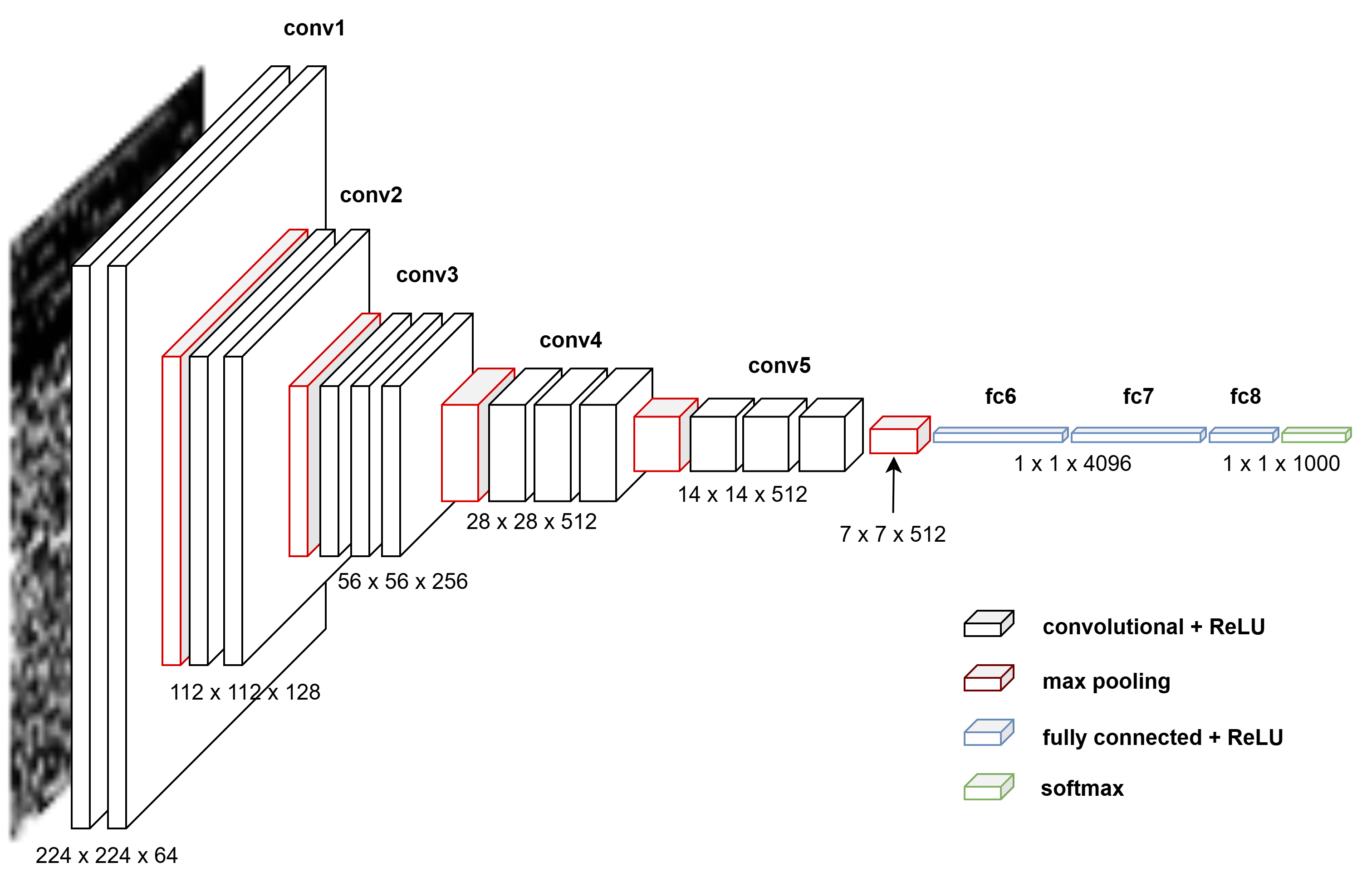}
\caption{Architecture of the VGG16 convolutional neural network.}%, showing the sequential convolutional, max-pooling, and fully connected layers used for feature extraction and classification of input Byte plot images.}
\label{fig:vgg16}
\end{figure}

\begin{figure}[t]%% placement 
\centering%% For centre alignment of image.
\includegraphics[scale=.10]{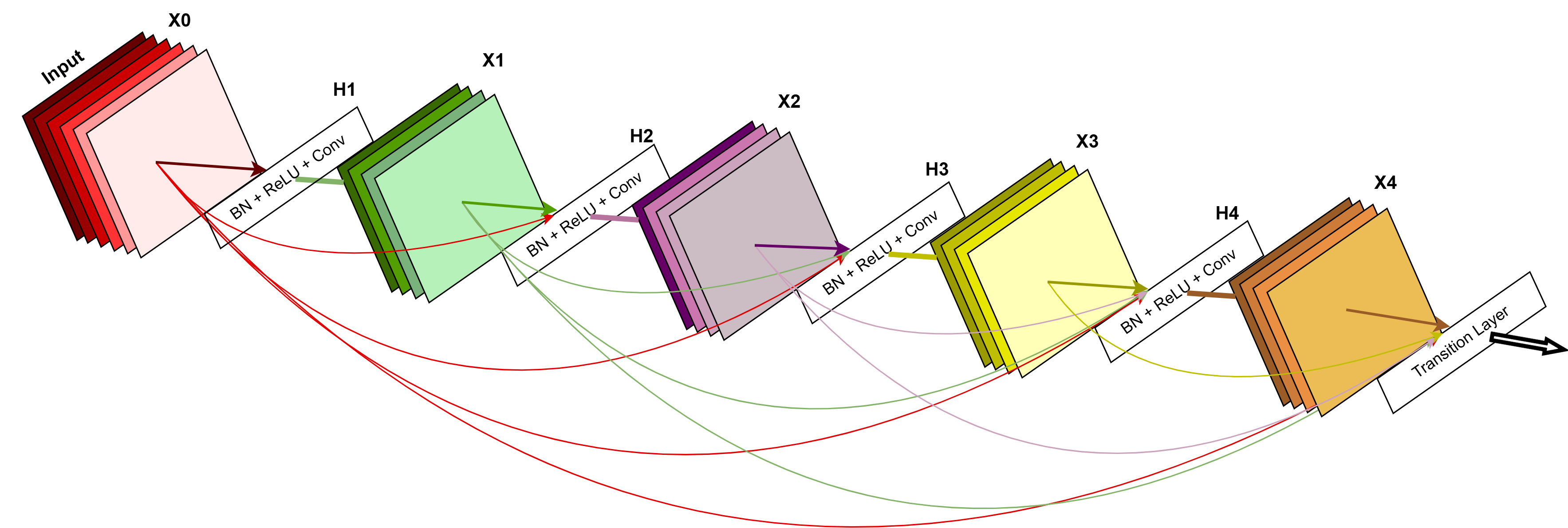}
\caption{Architecture of the DenseNet121 convolutional neural network.}% The diagram illustrates the sequence of densely connected blocks (X0–X4), where each layer receives feature maps from all preceding layers through direct skip connections. Each dense layer consists of Batch Normalization, ReLU activation, and a convolution operation, while transition layers reduce spatial resolution via convolution and pooling. This configuration enables efficient feature reuse and compact representations for classifying grayscale byte-plot malware images.}
\label{fig:densenet-121}
\end{figure}

We employed VGG16 and DenseNet121 because they offer strong, well-validated performance for detecting packed and non-packed executables. Both architectures were adapted using a transfer-learning paradigm, in which ImageNet-pretrained weights were retained for feature extraction, and a custom classification head was appended for binary classification. The overall architectures of the two CNNs, VGG16 and DenseNet121, are shown in Figures~\ref {fig:vgg16} and~\ref {fig:densenet-121}, respectively.

Input data consisted of grayscale byte-plot images of binary executables, resized to $224 \times 224$ pixels and expanded to three channels to match the requirements of the pretrained network. Pixel values were normalized using architecture-specific preprocessing functions. To mitigate class imbalance between “Packed” (label = 1) and “Non-Packed” (label = 0) samples, Random Over-Sampling (ROS) was applied. The resulting balanced dataset contained \textit{X} samples per class, partitioned into 70\% training, 15\% validation, and 15\% testing sets via stratified split.

VGG16 comprises 13 convolutional layers and three fully connected layers, with ReLU activation and max-pooling after each convolutional block. The pretrained convolutional base was frozen to preserve hierarchical spatial feature representations. A custom classification head was appended, comprising a dense layer (256 units, ReLU activation), dropout ($p=0.5$), and a sigmoid output unit for binary classification. The architecture is summarized in Table~\ref{tab:vgg16-arch}.

\begin{table}[htb]
\footnotesize
\centering
\caption{VGG16-Based Binary Classifier Architecture}
\label{tab:vgg16-arch}
\begin{tabular}{|l|c|r|c|}
\hline
\textbf{Layer (Type)} & \textbf{Output Shape} & \textbf{Param \#} & \textbf{Trainable} \\
\hline\hline
VGG16 (Functional) & (7, 7, 512) & 14,714,688 & No \\
Flatten & (25088) & 0 & No \\
Dense (ReLU, 256) & (256) & 6,422,784 & Yes \\
Dropout ($p=0.5$) & (256) & 0 & Yes \\
Dense (Sigmoid) & (1) & 257 & Yes \\
\hline
\multicolumn{2}{|r|}{Total parameters} & \multicolumn{2}{l|}{21,137,729} \\
\multicolumn{2}{|r|}{Trainable parameters} & \multicolumn{2}{l|}{6,423,041} \\
\multicolumn{2}{|r|}{Non-trainable parameters} & \multicolumn{2}{l|}{14,714,688} \\
\hline
\end{tabular}
\end{table}

DenseNet121 employs dense connectivity, in which each layer receives input from all preceding layers, thereby improving gradient flow and feature reuse while reducing the total number of parameters. Its convolutional base was frozen, and a custom head was appended, consisting of global average pooling, a dense layer with 256 units (ReLU), dropout ($p=0.5$), and a sigmoid output. The architecture is summarized in Table~\ref{tab:densenet121-arch}.

\begin{table}[!htb]
\footnotesize
\centering
\caption{DenseNet121-Based Binary Classifier Architecture}
\label{tab:densenet121-arch}
\begin{tabular}{|l|c|r|c|}
\hline
\textbf{Layer (Type)} & \textbf{Output Shape} & \textbf{Param \#} & \textbf{Trainable} \\
\hline\hline
DenseNet121 (Functional) & (7, 7, 1024) & 7,978,856 & No \\
GlobalAveragePooling2D & (1024) & 0 & No \\
Dense (ReLU, 256) & (256) & 262,400 & Yes \\
Dropout ($p=0.5$) & (256) & 0 & Yes \\
Dense (Sigmoid) & (1) & 257 & Yes \\
\hline
\multicolumn{2}{|r|}{Total parameters} & \multicolumn{2}{l|}{8,241,513} \\
\multicolumn{2}{|r|}{Trainable parameters} & \multicolumn{2}{l|}{262,657} \\
\multicolumn{2}{|r|}{Non-trainable parameters} & \multicolumn{2}{l|}{7,978,856} \\
\hline
\end{tabular}
\end{table}

Both models were trained using binary cross-entropy loss and the Adam optimizer with a learning rate of $\eta = 10^{-3}$ and a batch size of 32. Early stopping with a patience of 5 epochs was applied to prevent overfitting. ModelCheckpoint was used to save the best-performing weights. Multi-run experiments (5 runs) were conducted to compute mean performance metrics and 95\% confidence intervals for accuracy, precision, recall, and F1-score.

\section{Dataset description and motivation}\label{sec:dataset}
The dataset used in this study comprises a broad collection of packed and unpacked Portable Executable (PE) files spanning benign Win32, Win64, and .NET applications and malicious Win32 malware, ensuring wide coverage of structural and behavioral diversity observed in real-world environments. Table~\ref{tab:packers} illustrates the composition of the packed dataset and the extensive range of commercial and open-source packers applied to both benign and malicious executables, covering lightweight compression-based packers such as UPX as well as more advanced virtualization- and transformation-based packers including Themida, Enigma Virtual Box, PECompact, and related tools, each introducing distinct entropy patterns, section modifications, and header manipulations. Table~\ref{tab:nonpacked} presents the non-packed portion of the dataset, detailing benign software categories, general applications, Windows system files, and .NET executables, alongside malicious non-packed Windows malware. This inclusion of multiple executable architectures, runtime environments, and obfuscation profiles across packed and unpacked samples supports realistic evaluation of PE analysis and detection techniques. Benign software in both packed and unpacked forms was collected from Windows 11 system applications and from trusted software repositories such as CNET and Softpedia. In contrast, maliciously packed and unpacked samples were sourced from VirusTotal. Packing operations were performed either manually or via automated scripts, as supported by the respective packer.

To maintain experimental balance and generalizability, the packed samples were derived from both benign and malicious executables. However, in this context, the classifier is designed solely to distinguish between \textit{packed} and \textit{non-packed} binaries rather than to discriminate between malware and benignware. This design choice reflects a practical and security-relevant abstraction: in real-world analysis pipelines, packing itself constitutes a critical layer of obfuscation, independent of the binary's underlying intent. By training a convolutional neural network (CNN) to recognize packing artifacts, the model can effectively learn low-level statistical and spatial representations associated with packing transformations without bias toward specific malicious or benign characteristics.

This dataset, therefore, serves as a comprehensive benchmark for studying the impact of packing on PE structures and for validating CNN-based models’ ability to differentiate between packed and non-packed binaries. 
\begin{table*}
\footnotesize
\centering
\caption{Dataset Composition of Packed Executables}
\begin{tabular}{|l r|l r|}
\hline
\multicolumn{2}{|c|}{\textbf{Benign Packed}} & \multicolumn{2}{|c|}{\textbf{Malicious Packed}} \\ \hline
\textbf{Packer} & \textbf{Count} & \textbf{Packer} & \textbf{Count} \\
\hline\hline
ASPack & 514 & UPX & 870 \\
Alienyze & 126 & Aspack & 856 \\
Amber & 152 & AEGIS & 869 \\
BeRoEXEPacker & 116 & PECompact & 862 \\
EXpressor & 121 & NSPack & 869 \\
Enigma Virtual Box & 123 & MPRESS & 205 \\
\cline{3-4} % Line only on Malicious side, above the Malicious Total
% --- Malicious Total is placed here ---
Eronana Packer & 151 & \multicolumn{2}{r|}{\textbf{Total Malicious Packed 4531}} \\
\cline{3-4}
%\hline % <--- ADDED LINE HERE TO CLOSE THE DATA SECTION
Exe32pack & 128 & \multicolumn{2}{l|}{} \\
FSG & 119 & \multicolumn{2}{l|}{} \\
JDPack & 120 & \multicolumn{2}{l|}{} \\
MEW & 116 & \multicolumn{2}{l|}{} \\
MPRESS & 116 & \multicolumn{2}{l|}{} \\
Molebox & 119 & \multicolumn{2}{l|}{} \\
NSPack & 447 & \multicolumn{2}{l|}{} \\
Neolite & 116 & \multicolumn{2}{l|}{} \\
PECompact & 508 & \multicolumn{2}{l|}{} \\
PEtite & 440 & \multicolumn{2}{l|}{} \\
Packman & 115 & \multicolumn{2}{l|}{} \\
RLPack & 115 & \multicolumn{2}{l|}{} \\
TELock & 120 & \multicolumn{2}{l|}{} \\
Themida & 123 & \multicolumn{2}{l|}{} \\
UPX & 123 & \multicolumn{2}{l|}{} \\
WinUpack & 120 & \multicolumn{2}{l|}{} \\
Yoda-Crypter & 118 & \multicolumn{2}{l|}{} \\
Yoda-Protector & 115 & \multicolumn{2}{l|}{} \\
AEGIS & 408 & \multicolumn{2}{l|}{} \\
\cline{1-2} % Line separating Benign entries from Benign Total
\textbf{Total Benign Packed} & \textbf{4889} & \multicolumn{2}{l|}{} \\
\hline
\multicolumn{4}{|c|}{\textbf{Total Packed Executables: 9420}} \\
\hline
\end{tabular}
\label{tab:packers}
\end{table*}
\begin{table}[ht]
\footnotesize
\centering
\caption{Dataset Composition of Non-Packed Executables}
\begin{tabular}{|l r|l r|}
\hline
\multicolumn{2}{|c|}{\textbf{Benign Non-Packed}} & \multicolumn{2}{c|}{\textbf{Malicious Non-Packed}} \\ \hline
\textbf{Category} & \textbf{Count} & \textbf{Category} & \textbf{Count} \\
\hline\hline
General Applications & 428 & VirusTotal Win32 Malware & 4710 \\ \cline{3-4}
Windows System & 2136 &  \textbf{Total Malicious Non-Packed} & \textbf{4710}  \\ \cline{3-4} 
.NET Applications & 2146 &  &  \\  \cline{1-1} 
\cline{1-2}
\textbf{Total Benign Non-Packed} & \textbf{4710} &  &\\
\hline 
\multicolumn{4}{|c|}{\textbf{Total Non-Packed Executables: 9420}} \\
\hline
\end{tabular}
\label{tab:nonpacked}
\end{table}

From the perspective of antivirus (AV) and automated analysis systems, distinguishing between packed and non-packed executables is a \textit{fundamental prerequisite} rather than a secondary concern. When a binary is packed, its original code and data are transformed into an obfuscated format that conceals its proper functionality. Consequently, even advanced static or heuristic analysis engines cannot reliably determine whether a packed file is benign or malicious until it is successfully unpacked or emulated at runtime. This limitation renders the act of packing detection a critical stage in the security analysis pipeline. By identifying packed executables early, AV engines can selectively trigger unpacking routines, dynamic sandboxing, or emulation-based inspection before making a classification decision. Analysts often encounter sophisticated packers, such as those that devirtualize VM-protected binaries by translating custom virtual opcodes into readable machine instructions, thereby reconstructing the original program logic to understand behavior and extract indicators. This process is CPU- and memory-intensive, frequently relying on custom emulation or memory-dump tools, and complex samples often require long automated runs and manual reverse engineering. As a result, these tasks are typically performed on highly efficient, high-performance servers to provide the necessary compute resources, memory, and runtime stability, underscoring the critical role of packing detection in optimizing the overall analysis workflow. %% citation is needed here, maybe like this https://github.com/JonathanSalwan/VMProtect-devirtualization?tab=readme-ov-file

\section{Experimental setup and configuration}
\label{sec:experimental-setup}

This section describes the experimental environment, preprocessing pipeline, and training configuration used to evaluate both classical machine-learning models and deep convolutional neural networks (CNNs) for binary classification of packed versus non-packed executable files.

\subsection{Hardware and software environment}
All experiments were conducted on an Apple MacBook Pro powered by the Apple M1 system-on-chip (SoC), which features an 8-core CPU, an 8-core GPU, and 16 GB of unified memory. This hardware configuration provides balanced computational resources for both CPU-bound and GPU-accelerated deep-learning tasks.

The software environment was based on TensorFlow with its high-level Keras API. GPU acceleration on Apple Silicon was enabled through Apple’s \texttt{tensorflow-metal} plugin. NumPy and Scikit-learn were used for numerical computation, dataset preparation, and metric evaluation. Image preprocessing (loading, resizing, normalization, and augmentation) was performed using the \texttt{tensorflow.keras.preprocessing.image} module. To mitigate class imbalance in the classical ML experiments, RandomOverSampler from the \texttt{imblearn.over\_sampling} package was applied.

For Gabor jet feature extraction, OpenCV (\texttt{cv2}) was used to construct multi-scale, multi-orientation Gabor kernels and apply convolution-based image filtering. NumPy was employed to aggregate the mean and variance of each Gabor response into fixed-length feature vectors, which were subsequently used to train classical machine-learning models implemented in Scikit-learn.
\subsection{Results}

Evaluation of classical machine-learning models using Gabor jet features revealed clear performance stratification across the tested algorithms. Ensemble-based classifiers demonstrated a significant advantage over their non-ensemble counterparts. Random Forest achieved the strongest and most stable results, with accuracy, precision, recall, and F1-score converging to approximately 0.948 across the five runs. Its false positive rate (FPR) averaged 0.0361, substantially lower than those of other classical models, indicating that it rarely misclassified benign (non-packed) files as packed. XGBoost showed the second-best performance with accuracy, precision, recall, and F1-score all around 0.937. Although slightly less stable than Random Forest, it consistently outperformed the remaining algorithms.

KNN achieved moderate performance, with an average accuracy of 0.916, but its notably higher FPR (0.0873) indicates a greater tendency to misclassify non-packed samples. SVM and Logistic Regression, which rely on linear or margin-based decision boundaries, exhibited substantial performance degradation with accuracies of 0.826 and 0.814, respectively. These results indicate that linear models struggle to capture the complexity of the feature space generated by Gabor jets. The MLP model displayed the weakest and most unstable behavior, with particularly high variance across runs and an FNR of 0.120 in the worst-performing iteration. This instability suggests sensitivity to initialization and hyperparameters, making it less reliable for practical deployment, as illustrated in Table~\ref{tab:classification_results_final}.

\begin{sidewaystable*}
\footnotesize	
\centering
\begin{tabular}{|l|c|c|c|c|c|c|c|}
\hline
Model & Features & Precision & Recall &
F1-score & Accuracy &
FPR & FNR \\
\hline\hline

Random Forest & Gabor jets & $0.9483 \pm 0.0019$ & $0.9479 \pm 0.0019$ & $0.9478 \pm 0.0019$ & $0.9479 \pm 0.0019$ & $0.0361 \pm 0.0018$ & $0.0682 \pm 0.0020$ \\

XGBoost & Gabor jets & $0.9371 \pm 0.0026$ & $0.9368 \pm 0.0024$ & $0.9368 \pm 0.0024$ & $0.9368 \pm 0.0024$ & $0.0495 \pm 0.0056$ & $0.0769 \pm 0.0017$ \\

KNN & Gabor jets & $0.9161 \pm 0.0026$ & $0.9161 \pm 0.0027$ & $0.9161 \pm 0.0027$ & $0.9161 \pm 0.0027$ & $0.0873 \pm 0.0054$ & $0.0806 \pm 0.0008$ \\

SVM & Gabor jets & $0.8287 \pm 0.0013$ & $0.8256 \pm 0.0012$ & $0.8252 \pm 0.0012$ & $0.8256 \pm 0.0012$ & $0.2214 \pm 0.0031$ & $0.1275 \pm 0.0018$ \\

Logistic Regression & Gabor jets & $0.8162 \pm 0.0019$ & $0.8142 \pm 0.0020$ & $0.8139 \pm 0.0020$ & $0.8142 \pm 0.0020$ & $0.2271 \pm 0.0033$ & $0.1446 \pm 0.0019$ \\

MLP & Gabor jets & $0.8166 \pm 0.0348$ & $0.7757 \pm 0.1040$ & $0.7616 \pm 0.1283$ & $0.7757 \pm 0.1040$ & $0.3286 \pm 0.3009$ & $0.1200 \pm 0.0946$ \\
\hline
DenseNet121 & Imgs &
$0.9797 \pm 0.0050$ &
$0.9425 \pm 0.0088$ &
$0.9607 \pm 0.0042$ &
$0.9615 \pm 0.0040$ &
$0.0195 \pm 0.0050$ &
$0.0575 \pm 0.0088$ \\

VGG16 & Imgs &
$0.9771 \pm 0.0064$ &
$0.9523 \pm 0.0061$ &
$0.9645 \pm 0.0027$ &
$0.9650 \pm 0.0027$ &
$0.0224 \pm 0.0064$ &
$0.0477 \pm 0.0061$ \\
\hline
\end{tabular}
\caption{Performance metrics for classical and CNN-based models in detecting packed and non-packed executables. Values represent mean ($\mu$) and standard deviation ($\sigma$) over five independent runs. DenseNet121 achieves the highest precision and lowest FPR, while VGG16 achieves the highest recall and lowest FNR.}
\label{tab:classification_results_final}
\end{sidewaystable*}

\begin{figure}[h!]
    \centering
    \includegraphics[width=1\textwidth]{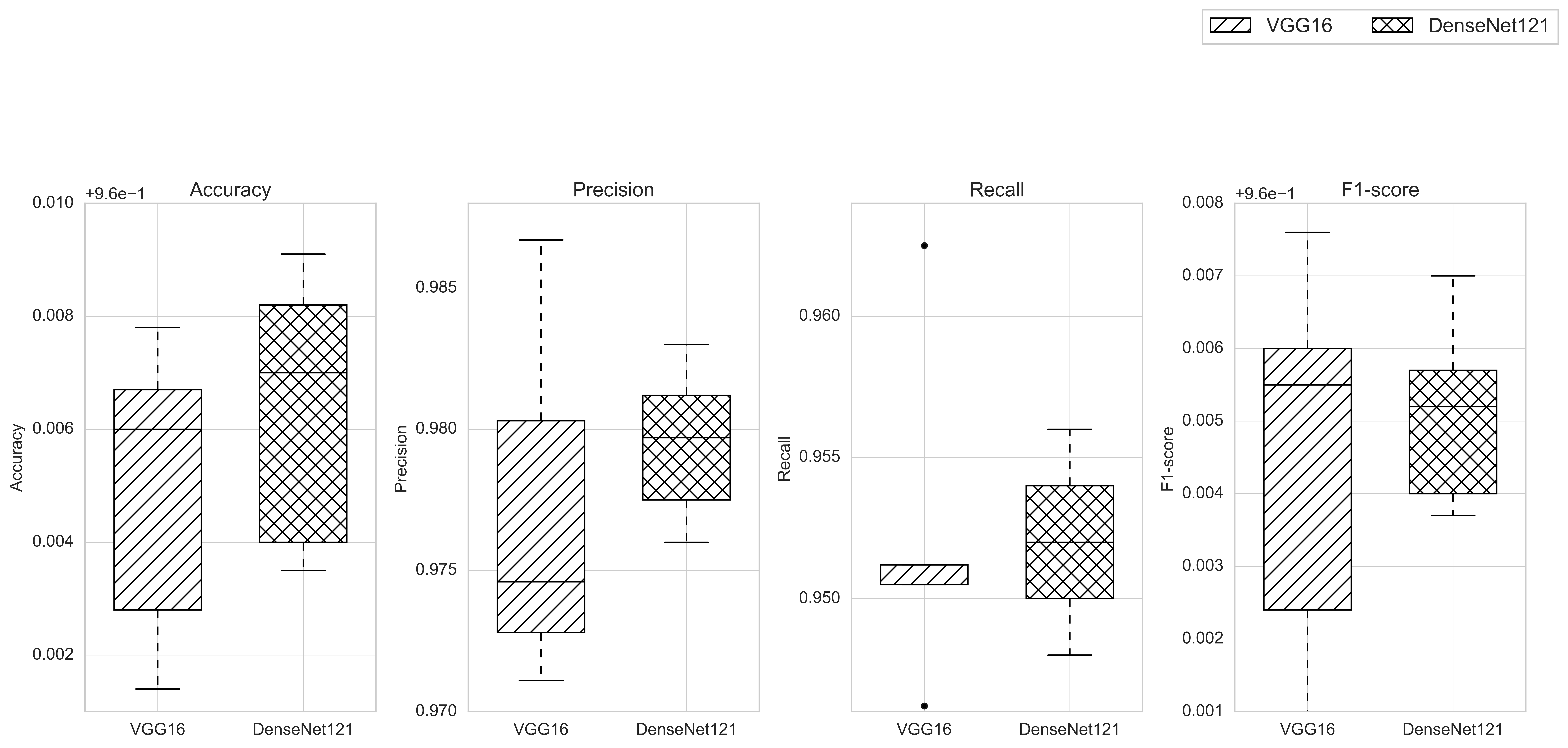}
    \caption{Boxplots of Accuracy, Precision, Recall, and F1-score across five runs for VGG16 and DenseNet121. Whiskers indicate the range of observed values, and hatch patterns distinguish the models, highlighting consistent performance with low variability.}
    \label{fig:metrics_boxplot}
\end{figure}

In contrast to the classical models, the convolutional neural networks (CNNs) demonstrated markedly superior performance across all metrics. Both DenseNet121 and VGG16 were trained with early stopping, with patience set to 5 epochs. DenseNet121 converged after 30 epochs, while VGG16 converged after 10 epochs. The training and validation curves for both models indicated stable convergence and nearly parallel trajectories between training and validation performance, suggesting strong generalization with minimal overfitting. DenseNet121 achieved the highest precision among all evaluated models, averaging 0.9797 with a corresponding FPR of only 0.0195. This behavior reflects a conservative decision boundary that minimizes false alarms, an important property when misclassifying a clean file as packed may trigger unnecessary unpacking, sandboxing, or static analysis overhead. VGG16, on the other hand, achieved the highest recall (0.9523) and lowest FNR (0.0477), indicating stronger sensitivity in detecting packed executables. This makes VGG16 particularly suitable for scenarios in which missing a packed (and potentially malicious) file is more costly than issuing a false alarm.

Both CNNs achieved higher F1-scores and overall accuracy than all classical models, highlighting the advantage of automatically learned image-based representations over hand-crafted Gabor features. DenseNet121 averaged 0.9615 accuracy and 0.9607 F1-score, while VGG16 achieved slightly higher values (0.9650 accuracy and 0.9645 F1-score). These results underline the robustness of deep feature extraction and the models’ ability to capture complex, high-dimensional patterns in packed and non-packed binary visualizations. Figure~\ref{fig:metrics_boxplot} illustrates the boxplots of Accuracy, Precision, Recall, and F1-score across five runs for VGG16 and DenseNet121. The metrics for both models are very close, with only slight differences observed, and the compact boxplots indicate low variability across runs.

\begin{figure}[h!]
    \centering
    \includegraphics[width=0.5\textwidth]{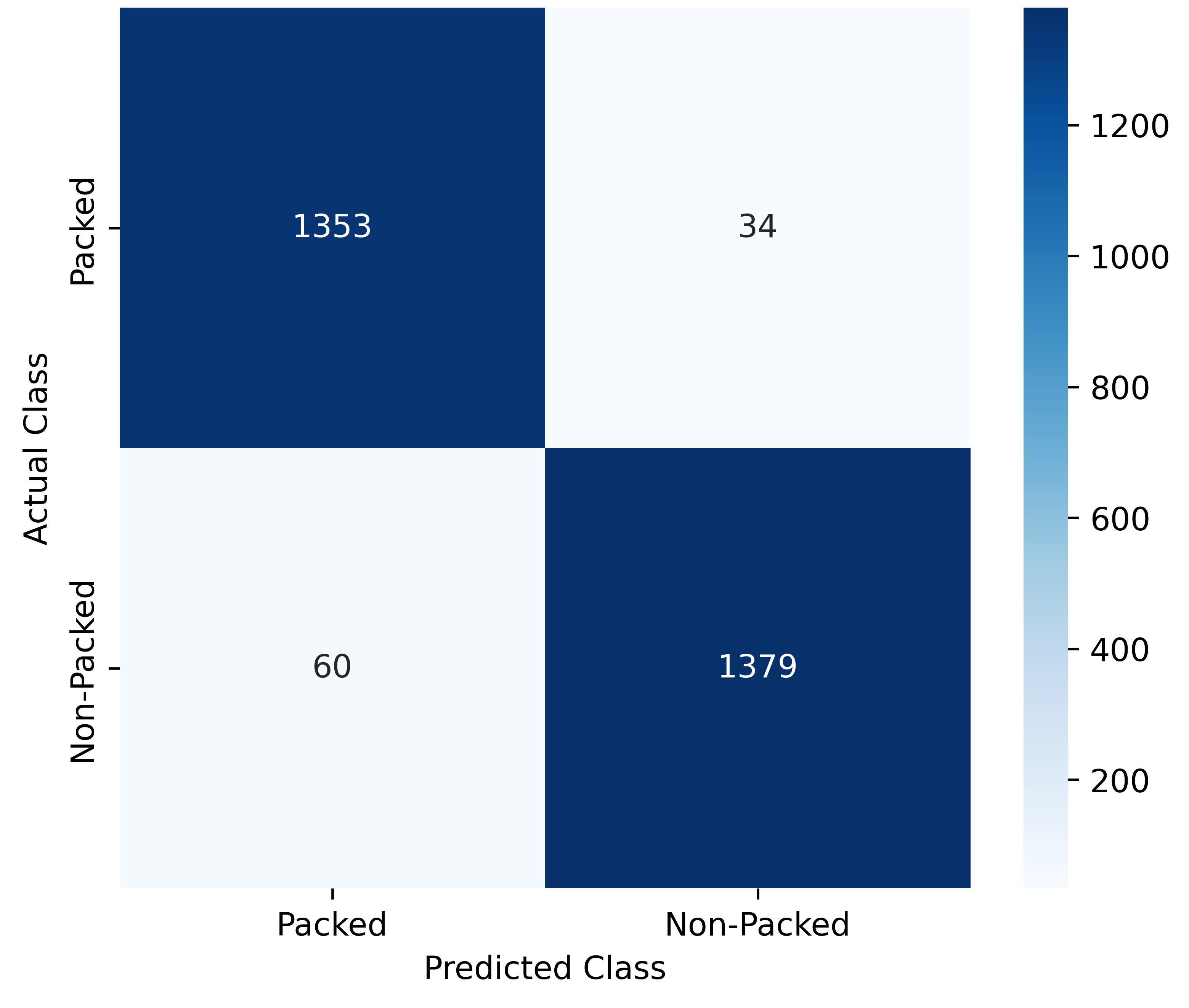}
    \caption{VGG16 Confusion Matrix (Run 3) showing predicted vs actual classes.}
    \label{fig:vgg16_run5_confusion}
\end{figure}

\begin{figure}[h!]
    \centering
    \includegraphics[width=0.5\textwidth]{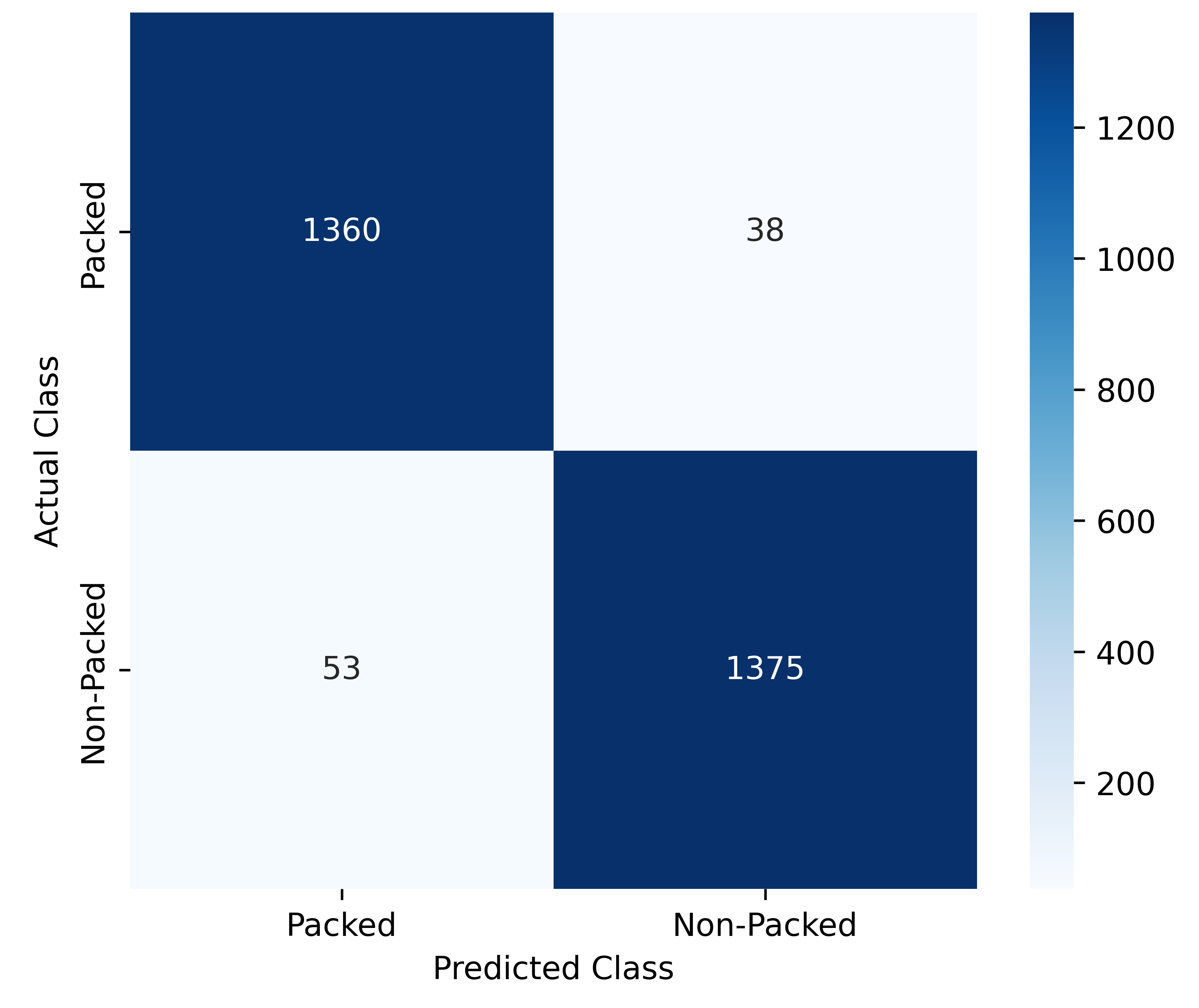}
    \caption{DenseNet121 Confusion Matrix (Run 5) showing predicted vs actual classes.}
    \label{fig:densenet121_run5_confusion}
\end{figure}

The best-performing runs for each model were Run 3 for VGG16 and Run 5 for DenseNet121. In Run 3 of the VGG16 model, as illustrated in Figure~\ref{fig:vgg16_run5_confusion}, the confusion matrix showed 1353 true positives (TP), 34 false positives (FP), 1379 true negatives (TN), and 60 false negatives (FN), resulting in an accuracy of 96.7\%, precision of 97.5\%, recall of 95.7\%, and an F1-score of 96.4\%. As shown in Figure~\ref{fig:densenet121_run5_confusion} for DenseNet121 in Run 5, the confusion matrix revealed 1360 TP, 38 FP, 1375 TN, and 53 FN, with an accuracy of 96.8\%, precision of 97.3\%, recall of 96.2\%, and an F1-score of 96.8\%. While both models performed exceptionally well, DenseNet121 achieved slightly higher recall and F1-score (96.8\%) than VGG16, which had somewhat lower recall but higher precision.

Overall, the results demonstrate that CNN-based approaches clearly outperform classical machine-learning techniques for the task of packed executable detection. DenseNet121 excels at reducing false positives, while VGG16 excels at reducing false negatives. Depending on the operational priorities of a malware analysis pipeline, either architecture can be deployed effectively to minimize unnecessary processing or missed packed threats. The close alignment of training and validation metrics further confirms that both models generalize well and are suitable for real-world deployment in large-scale malware triage systems.

%DenseNet121 was trained for up to 30 epochs with early stopping. Averaged training curves show training accuracy increasing from 85.4\% to 96.5\%, while validation accuracy improved from 89.9\% to 96.2\%, peaking between epochs 24--28. Training loss decreased from 0.348 to 0.097, and validation loss from 0.251 to 0.111. F1-scores exhibited similar growth patterns. Aggregated over five runs, DenseNet121 achieved an average accuracy of 96.15\%, precision of 97.97\%, recall of 94.25\%, and F1-score of 96.07\%, demonstrating strong predictive capability and no indication of overfitting.

\subsection{Model performance against unknown packers}
In recent years, the cybercrime ecosystem has witnessed the emergence of \emph{packers-as-a-service (PaaS)}, such as HeartCrypt and BatCloak, which provide on-demand payload obfuscation through web-based interfaces rather than standalone applications~\cite{tujague2024cryptedhearts}. This model allows operators to retain full control of their packing engines, continuously update evasion techniques, and monetize access through subscription or pay-per-use plans. The PaaS model not only lowers the barrier to entry for less-experienced threat actors but also accelerates the proliferation of highly evasive variants by abstracting away the technical complexity of packer development and maintenance.

Although our dataset includes a broad range of commercial and open-source packers, real-world malware frequently employs \emph{active or custom packers} that evolve rapidly to evade detection. These packers are often designed for specific campaigns, incorporating multilayer encryption, staged runtime unpacking, and anti-analysis techniques that challenge traditional static detection workflows. Advanced Persistent Threat (APT) groups are known to rely on such bespoke or semi-custom packers to protect their payloads, ensuring long-term stealth and operational resilience. These APT-oriented packers often integrate virtualization-based obfuscation, polymorphic stubs, environmental checks, and decryption routines that only execute under controlled runtime conditions, features that significantly complicate both machine-learning-based detection and post-intrusion forensic acquisition.

To evaluate the generalization capability of our best-performing models under realistic adversarial conditions, we constructed a dedicated test set comprising 3,661 benign non-packed and 2,866 benign packed samples, incorporating recent active or custom packing techniques. A key component of this evaluation is that all samples were generated using a modified version of \emph{Lime Crypter}~\cite{NYANxCAT_LimeCrypter_2019}, a modern crypter widely adopted by cybercriminals to evade static detection. Lime Crypter handles both native and .NET binaries, encrypting payloads within a .NET stub that conceals the original executable until runtime. Upon execution, the stub decrypts and loads the payload directly into memory, closely emulating the behavior of an advanced runtime packer. Although commonly referred to as a crypter, Lime Crypter functions as both a crypter and a packer through its encrypted encapsulation and live unpacking routines.

\begin{figure}
\centering
\includegraphics[scale=0.38]{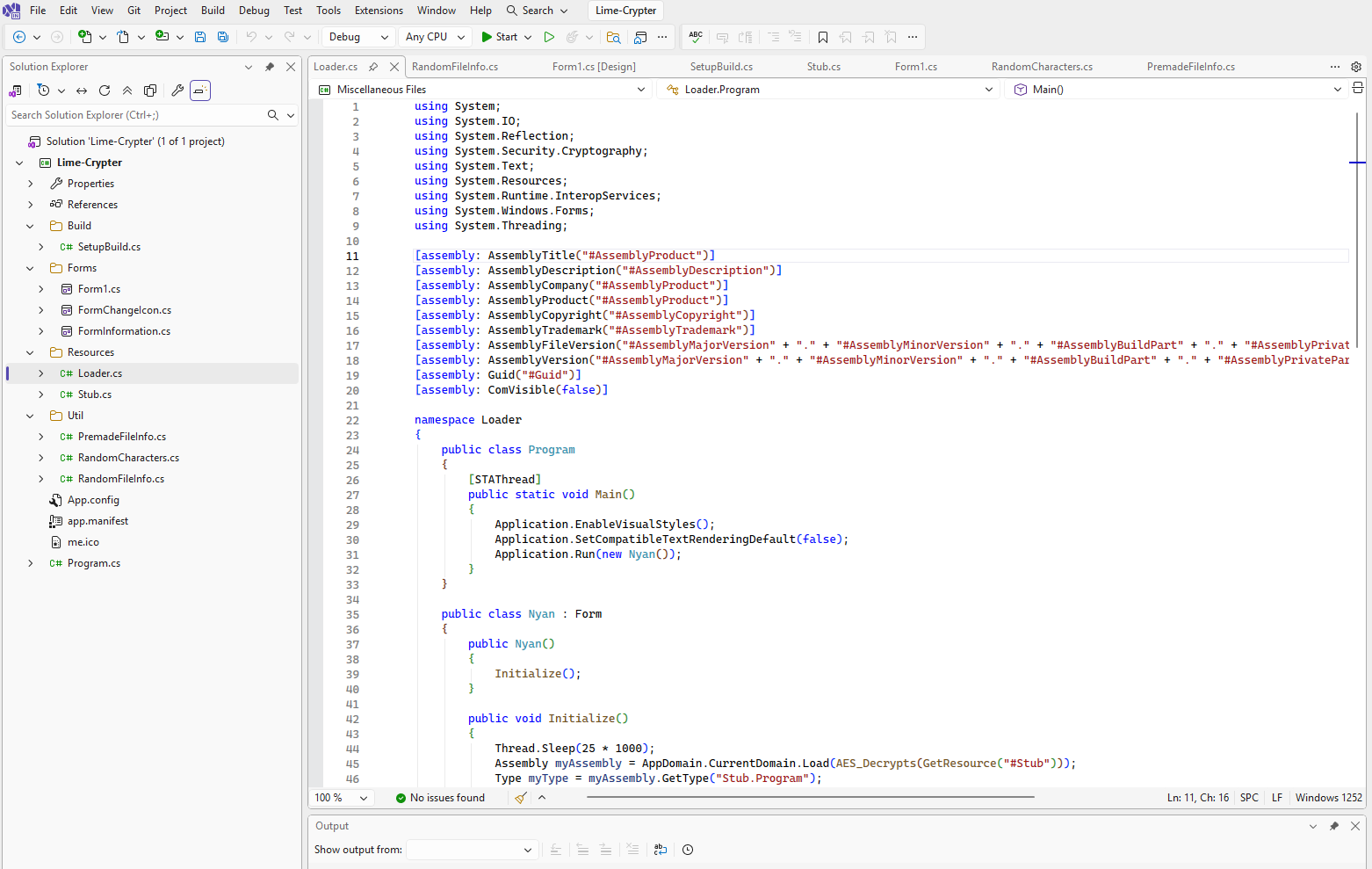}
\caption{Lime Crypter C\# source code.}
\label{fig:limecs_gui}
\end{figure}

For this experiment, we introduced additional adversarial variations by applying lightweight code-level modifications to the Lime Crypter project—such as renaming functions and classes and adding a compression stage before encryption. These adjustments mirror the subtle changes often made by malware authors to evade signature-based detection. Figure~\ref{fig:limecs_gui} shows the Visual Studio Project Explorer for the modified Lime Crypter project, providing a high-level view of its structure within our evaluation pipeline.

\begin{figure}
\centering
\includegraphics[scale=0.25]{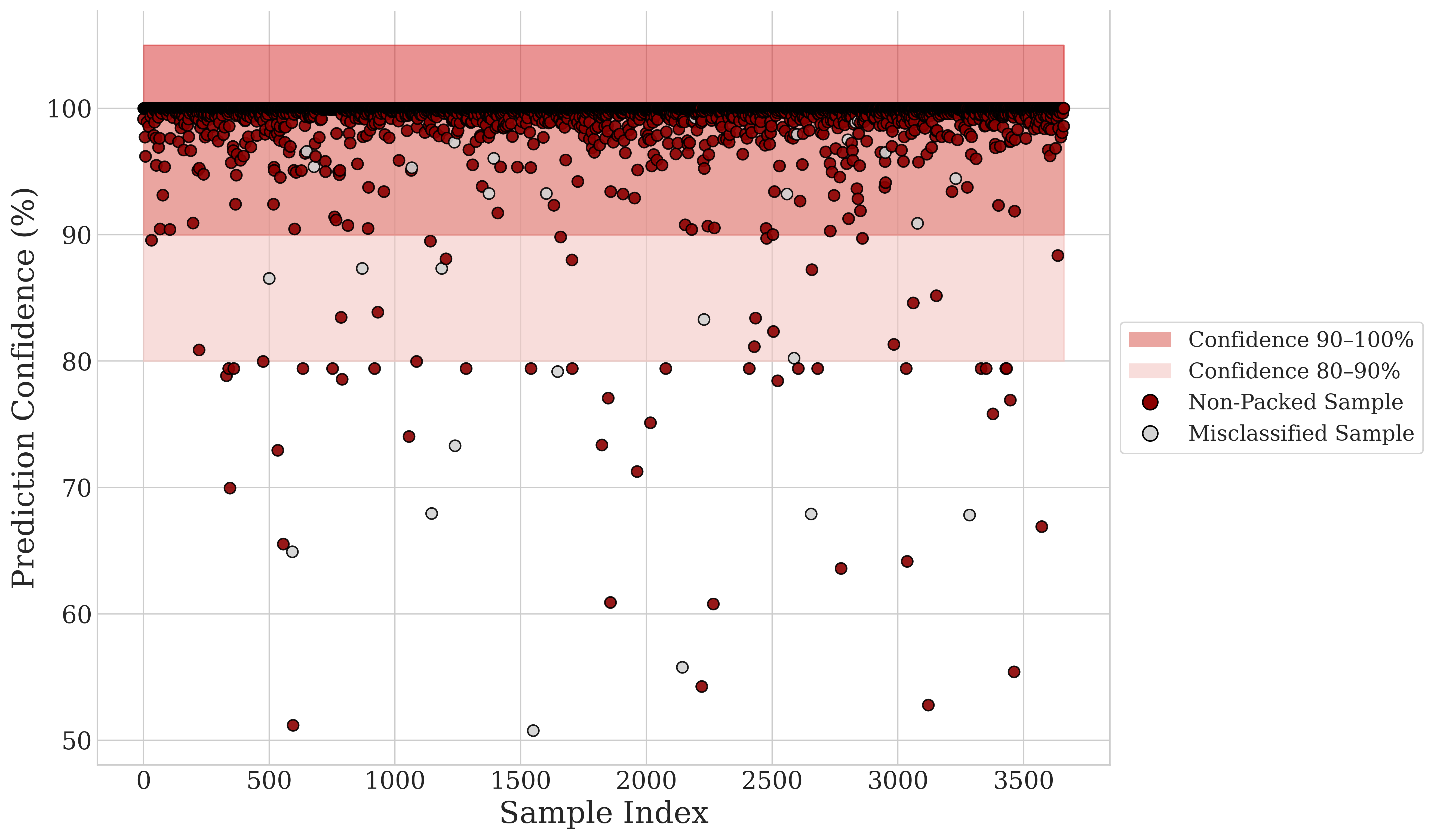}
\caption{VGG16 predictions for non-packed .NET samples, showing powerful performance with near-perfect confidence across all classifications.}
\label{fig:vgg_nonpacked}
\end{figure}

\begin{figure}
\centering
\includegraphics[scale=0.25]{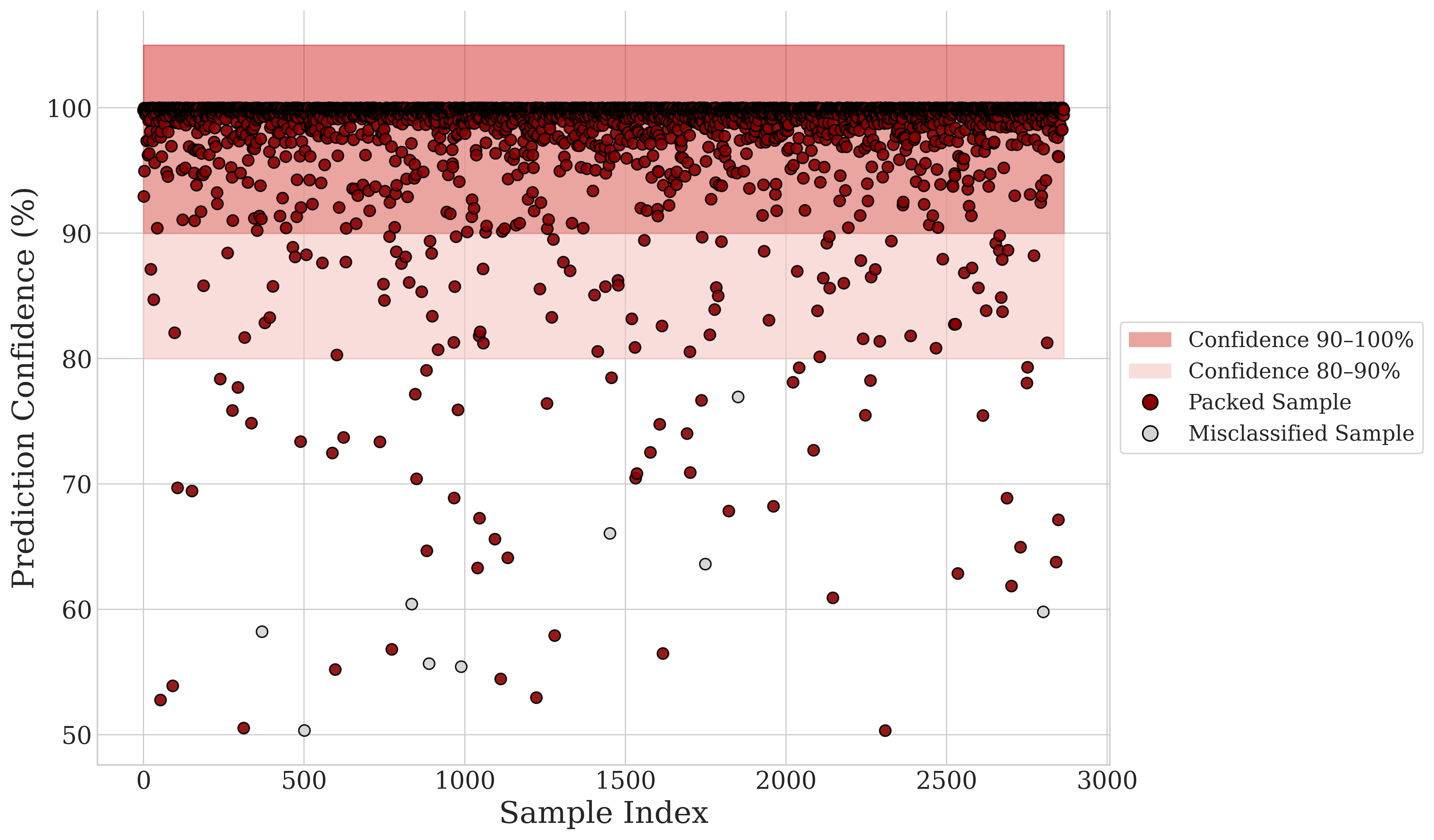}
\caption{VGG16 classification results on Lime Crypter-packed .NET binaries. The model identifies most packed samples, with only a small number misclassified as non-packed (false negatives).}
\label{fig:vgg_packed}
\end{figure}
Using this adversarial dataset, we evaluated the robustness of the two deep-learning models, VGG16 and DenseNet121. Both models had been thoroughly trained in prior experiments. For this evaluation, we used the model from a specific trained run for each CNN, relying solely on their exported .keras files. Testing these pre-trained architectures on identically packed samples enabled a direct and fair comparison of their resilience to obfuscation and runtime unpacking. This setup allowed us to precisely quantify how modern packing techniques influence model confidence, prediction stability, and overall detection reliability across different convolutional backbones.

To further validate the impact of these alterations, we analyzed the samples using \emph{Detect It Easy} (DIE)~\citep{DetectItEasy}, a widely used signature-based packer identification tool. Before modification, DIE consistently recognized original Lime Crypter-packed binaries by leveraging function-name and structural signatures embedded within the stub. However, after applying our adversarial modifications, the detector failed to identify the packer in all cases. These results demonstrate the inherent fragility of signature-based approaches: even superficial changes, such as renaming functions or altering the sequence of compression and encryption, are sufficient to invalidate existing rules and render detection ineffective. This underscores the limitations of static, signature-driven workflows in modern threat landscapes where packers evolve rapidly to evade such heuristics.

To further examine prediction behavior, we analyzed the \emph{confidence score distributions} produced by VGG16 for both packed and non-packed samples. Figure~\ref{fig:vgg_nonpacked} visualizes these distributions by plotting the confidence assigned to every individual prediction. This representation clearly illustrates how packing affects model certainty and highlights borderline cases in which the model struggles to distinguish adversarially packed malware from benign packed binaries. Such insights are crucial for understanding detection robustness in environments where actively maintained or custom packers, such as Lime Crypter, are routinely leveraged to evade static machine learning systems.
Figure~\ref{fig:vgg_nonpacked} shows the model’s confidence across all non-packed samples. Correctly classified samples are marked in deep red, whereas misclassified samples are shown as pale gray circles. Background shading indicates confidence tiers: light pink for 80–90\% and mid-pink for 90–100\%. This plot demonstrates that VGG16 performs exceptionally well on non-packed samples. The results in Figure~\ref{fig:vgg_packed} show a modest reduction in confidence and a slight increase in false negatives when encountering Lime Crypter-packed binaries. Nevertheless, VGG16 maintains a tightly clustered high-confidence region and demonstrates strong predictive stability. This indicates that while packing disrupts feature representations, VGG16 remains comparatively robust under adversarial obfuscation.

\begin{figure}
\centering
\includegraphics[scale=0.25]{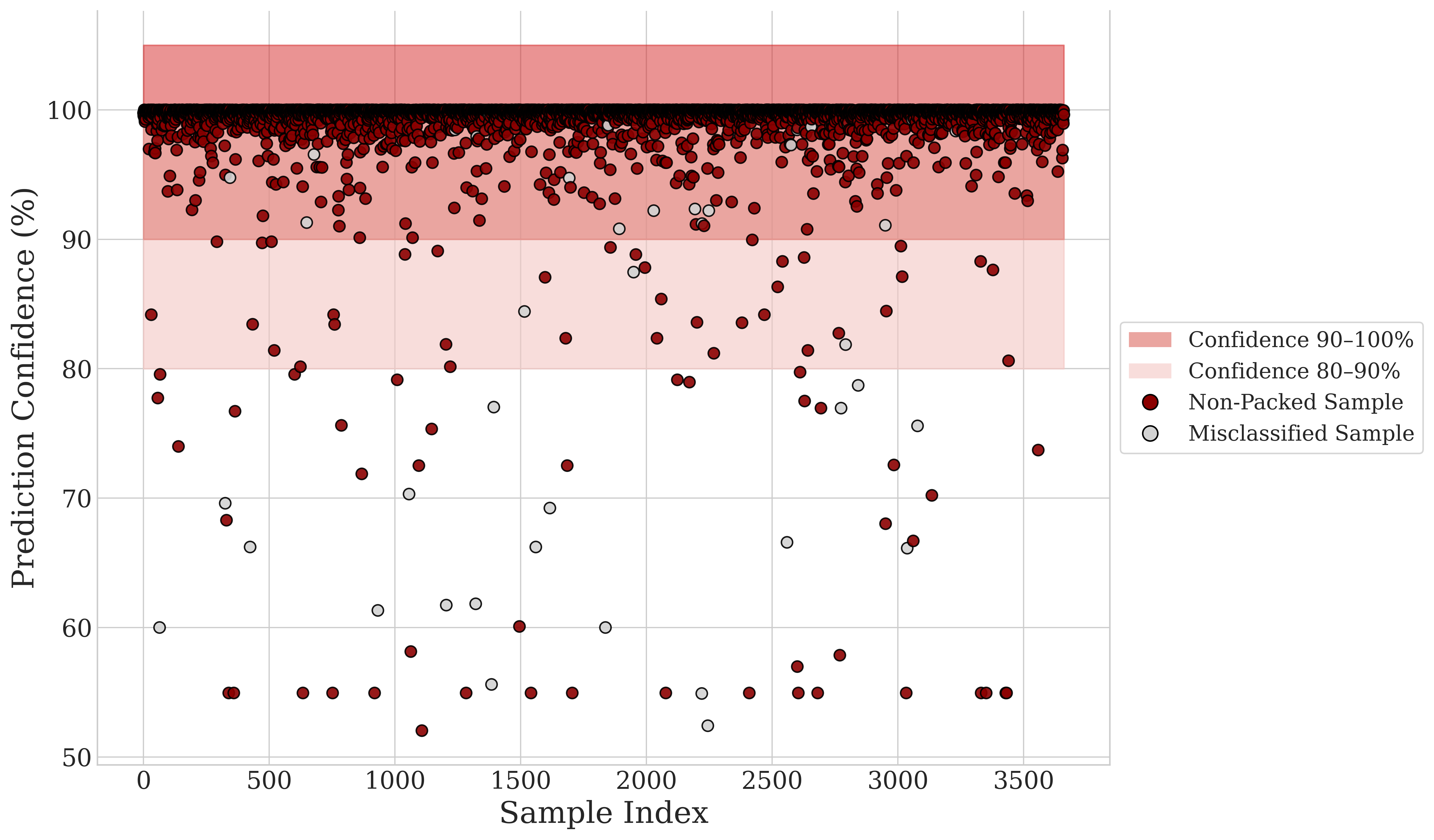}
\caption{DenseNet121 classification results on Lime Crypter-packed .NET binaries. While the model correctly identifies many packed samples, confidence scores are more dispersed, with a higher proportion of moderate-confidence predictions.}
\label{fig:dense2}
\end{figure}
\begin{figure}
\centering
\includegraphics[scale=0.25]{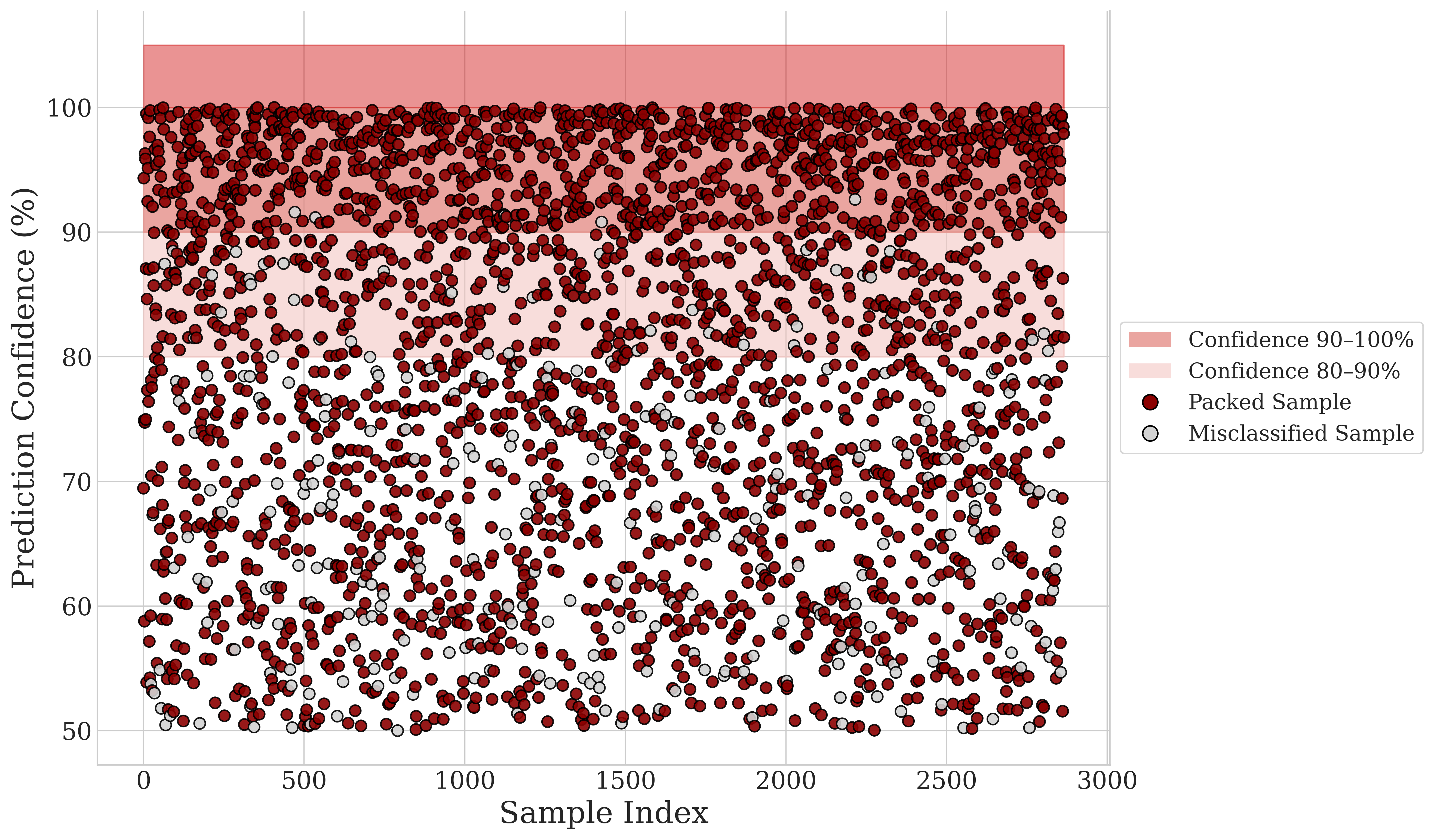}
\caption{DenseNet121 prediction confidence visualization for Lime Crypter-packed .NET binaries. The distribution is noticeably more scattered than VGG16, indicating reduced robustness under adversarial packing conditions.}
\label{fig:dense1}
\end{figure}

Although DenseNet121 performs reasonably well, its predictive behavior (Figures~\ref{fig:dense2} and \ref{fig:dense1}) shows greater variance than that of VGG16 under Lime Crypter packing. In Figure~\ref{fig:dense2}, predictions span a wider confidence range, with many samples falling between 70–90\%. Misclassifications are more frequent and do not exhibit the clean separation observed in the VGG16 plots.
Figure~\ref{fig:dense1} reinforces this observation: the confidence distribution forms a dense, highly scattered cloud across the 60–95\% range. Unlike VGG16, which retains a strong high-confidence band, DenseNet121 exhibits substantial dispersion, indicating that Lime Crypter’s obfuscation layers more strongly disrupt its learned features. Lower-confidence predictions occur more frequently, and the model exhibits reduced stability in distinguishing between packed and non-packed binaries.

Taken together, these results demonstrate that while DenseNet121 can detect Lime Crypter-packed binaries, VGG16 exhibits stronger generalization, higher-confidence predictions, and fewer false negatives under the same adversarial conditions. The tighter clustering and reduced variance in VGG16’s confidence distributions suggest that its representations are less susceptible to the distortions introduced by modern packing techniques, making it a more reliable classifier for detecting heavily obfuscated .NET binaries in this evaluation.

\section{Discussion}\label{sec:discussion}
The experimental results demonstrate the effectiveness of image-based representations for distinguishing packed from non-packed executables. By converting binary files into visual formats, such as byte plots, the models can capture structural and statistical patterns introduced by packing transformations. These patterns are often challenging to detect using conventional non-image features. The conversion process itself serves as a powerful preprocessing step, as it encodes complex byte-level dependencies and structural anomalies in a format that convolutional neural networks (CNNs) can readily exploit.

Among the CNN architectures evaluated, VGG16 and DenseNet121 exhibited complementary strengths. VGG16 achieved higher recall and lower false negative rates, indicating that it captures subtle packing patterns and reduces the likelihood of missing packed files. DenseNet121, on the other hand, achieved higher precision and lower false-positive rates, suggesting that its dense connectivity and feature reuse enable more conservative classification, thereby minimizing misclassification of benign executables. This distinction highlights the practical importance of model selection: for security-critical environments where false negatives incur high risk, VGG16 may be preferable, whereas DenseNet121 offers stronger reliability in reducing false alarms.

Beyond the controlled experimental dataset, the evaluation against \emph{unknown} or adversarial packers, specifically, samples packed with a modified version of Lime Crypter, provides critical insight into model robustness under realistic threat conditions. To mirror the subtle yet impactful alterations that malware authors commonly introduce to evade signature-based detection, we applied lightweight \emph{code-level} modifications to Lime Crypter. These changes included renaming functions and variables, introducing an additional compression stage prior to encryption, and modifying small portions of internal logic to alter the execution flow. Importantly, the overall Visual Studio project structure was left unchanged; the modifications targeted only the source-code internals, where signature-based heuristics typically anchor their detection rules. By altering the code without restructuring the project, we simulate realistic adversarial behaviors, such as identifier obfuscation, stub edits, and minor packing-routine changes, that routinely enable malware to evade static fingerprinting while preserving functional behavior.

The comparative results on this adversarial dataset reveal a notable divergence in the robustness of the two CNN models. VGG16 maintained tightly clustered, high-confidence predictions and exhibited lower variance when analyzing Lime Crypter-packed binaries, resulting in fewer false negatives and comparatively stable classification behavior. DenseNet121, while still capable of detecting many packed samples, showed noticeably wider confidence dispersion and a higher proportion of borderline predictions in the 60--90\% confidence range. These findings indicate that while DenseNet121 performs strongly on clean and broadly representative datasets, VGG16 generalizes more reliably when confronted with actively maintained or lightly modified packers that introduce structural shifts not present in the training data.

Compared with classical machine-learning methods built on handcrafted features, such as Gabor jets, CNN-based approaches achieved substantially higher performance across all primary metrics, particularly in generalizing to previously unseen packing techniques. Handcrafted features, while computationally efficient and interpretable, remain constrained by manually engineered patterns that often fail under adversarial obfuscation. Deep learning models, by contrast, automatically learn hierarchical spatial representations that capture both global structural trends and fine-grained byte-level anomalies, providing greater resilience against custom packing strategies.

Our findings underscore the importance of dataset quality and diversity in achieving reliable detection performance. A dataset encompassing commercial, open-source, and adversarial packing techniques exposed the models to realistic transformations, multilayer encryption, runtime unpacking behaviors, and obfuscation strategies typical of contemporary malware. Without such diversity, models risk overfitting to narrow, signature-like artifacts that limit practical applicability and degrade real-world robustness.

Finally, these results highlight promising directions for future research. While byte-plot images proved effective, alternative visual layers, such as multi-channel entropy maps, grayscale distribution histograms, or hybrid representations combining static and structural signals, may further enhance discriminative power. Likewise, evaluating more recent or lightweight CNN architectures optimized for real-time operation may improve the deployability of image-based packer detection. Overall, our study demonstrates that combining high-quality, diverse datasets with carefully selected CNN architectures provides a robust and practical solution for detecting packed binaries, reinforcing the value of image-based deep learning approaches in modern malware analysis and defense.

\section{Conclusion}\label{sec:conc}
This study demonstrates the effectiveness of image-based representations, specifically Byte plots, combined with deep learning models for detecting packed executables. By transforming binary files into grayscale images, the approach captures structural and statistical patterns introduced by a wide range of packing techniques, enabling robust discrimination between packed and non-packed binaries. Across all experiments, convolutional neural networks (CNNs), particularly VGG16 and DenseNet121, consistently outperformed classical feature-based baselines. The models achieved up to 96.5\% test accuracy with balanced precision, recall, and F1 Scores, confirming the suitability of CNN architectures for learning discriminative spatial features directly from byte-plot images. While handcrafted Gabor jet features remain computationally lightweight and functional in constrained environments, their discriminative capacity is limited when compared to learned deep representations. A key contribution of this work is the evaluation of model generalization against \emph{unknown or adversarial packers}. Using an actively modified version of Lime Crypter, featuring code-level alterations intended to mimic realistic evasion techniques, we showed that CNN-based methods retain strong predictive performance even when confronted with packers not seen during training. In particular, VGG16 exhibited higher-confidence predictions and greater resilience to adversarial obfuscation, underscoring the importance of architectural choice in modern malware detection. The findings highlight several practical implications for cybersecurity workflows. Early identification of packed binaries is essential, as packing often precedes more advanced obfuscation and runtime evasions. Reliable detection enables downstream tasks such as automated unpacking, behavioral inspection, and malware family attribution. More broadly, the combination of diverse datasets, realistic adversarial samples, and visually driven deep learning methods promotes stronger generalization in rapidly evolving threat landscapes. Future work may explore enhanced visual encodings, multi-channel representations, lightweight CNN architectures suitable for endpoint deployment, and hybrid approaches that fuse handcrafted and learned features.

\section*{CRediT authorship contribution statement}
Ehab Alkhateeb: Conceptualization, Methodology, Software, Investigation, Formal Analysis, Data Curation, Validation, Visualization, Writing – Original Draft; Ali Ghorbani: Conceptualization, Supervision; Arash Habibi Lashkari: Conceptualization, Writing – Review \& Editing, Supervision.

\section*{Conflict of interest} The authors declare that they have no known competing
financial interests or personal relationships that could appear to
influence the work reported in this paper.

\section*{Acknowledgments}{ The authors thank VirusTotal for providing the malware dataset used in this research.}

\bibliographystyle{unsrt}
\bibliography{REF}  %%% Uncomment this line and comment out the ``thebibliography'' section below to use the external .bib file (using bibtex) .

\clearpage
\appendix
\section{Appendix}

% ===================== FIGURE: VGG16 Accuracy (Corrected Averages, 10 Epochs) =====================
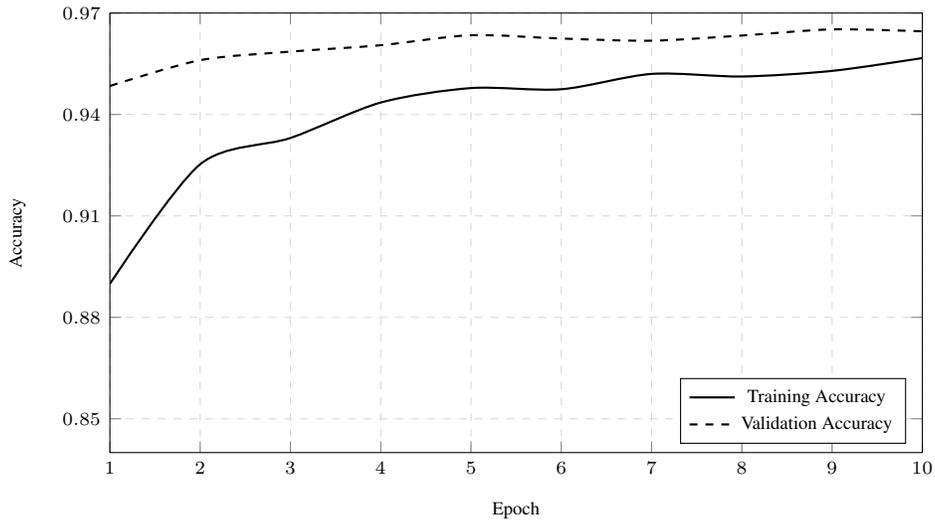
\begin{figure}[htbp]
\centering
\begin{tikzpicture}
\begin{axis}[
    width=0.75\textwidth,
    height=0.45\textwidth,
    xlabel={Epoch},
    ylabel={Accuracy},
    xmin=1, xmax=10,
    ymin=0.84, ymax=0.97,
    xtick={1,2,3,4,5,6,7,8,9,10},
    ytick={0.85,0.88,0.91,0.94,0.97},
    grid=major,
    grid style={dashed,gray!30},
    tick label style={font=\scriptsize},
    label style={font=\scriptsize},
    title style={font=\scriptsize, yshift=1ex},
    legend style={font=\scriptsize, at={(0.98,0.02)}, anchor=south east},
]

% === Training Accuracy (Averaged over first 10 epochs) ===
\addplot[color=black, thick, solid, smooth, line join=round]
coordinates {
(1,0.88991) (2,0.92528) (3,0.93305) (4,0.94353) (5,0.94781)
(6,0.94746) (7,0.95200) (8,0.95123) (9,0.95289) (10,0.95669)
};
\addlegendentry{Training Accuracy}

% === Validation Accuracy (Averaged over first 10 epochs) ===
\addplot[color=black, thick, dashed, smooth, line join=round]
coordinates {
(1,0.94843) (2,0.95606) (3,0.95859) (4,0.96048) (5,0.96344)
(6,0.96246) (7,0.96182) (8,0.96335) (9,0.96517) (10,0.96459)
};
\addlegendentry{Validation Accuracy}

\end{axis}
\end{tikzpicture}
\caption{VGG16 training and validation accuracy across 10 epochs (5-run average).}
\label{fig:vgg16-acc}
\end{figure}

% ===================== FIGURE 2: VGG16 Loss =====================
\begin{figure}[htbp]
\centering
\begin{tikzpicture}
\begin{axis}[
    width=0.75\textwidth,
    height=0.45\textwidth,
    xlabel={Epoch},
    ylabel={Loss},
    xmin=1, xmax=10,
    ymin=0.09, ymax=1.05,
    xtick={1,2,3,4,5,6,7,8,9,10},
    ytick={0.1,0.2,0.3,0.4,0.5,0.6,0.7,0.8,0.9,1.0},
    grid=major,
    grid style={dashed,gray!30},
    tick label style={font=\scriptsize},
    label style={font=\scriptsize},
    title style={font=\scriptsize, yshift=1ex},
    legend style={font=\scriptsize, at={(0.98,0.98)}, anchor=north east},
]
\addplot[color=black, thick, solid, smooth, line join=round] coordinates {
    (1,1.0039) (2,0.1917) (3,0.1751) (4,0.1423) (5,0.1417)
    (6,0.1360) (7,0.1245) (8,0.1162) (9,0.1117) (10,0.1063)
};
\addlegendentry{Training Loss}

\addplot[color=black, thick, dashed, smooth, line join=round] coordinates {
    (1,0.1537) (2,0.1336) (3,0.1323) (4,0.1141) (5,0.1248)
    (6,0.1160) (7,0.1211) (8,0.1272) (9,0.1166) (10,0.1271)
};
\addlegendentry{Validation Loss}
\end{axis}
\end{tikzpicture}
\caption{VGG16 training and validation loss across 10 epochs (5-run average).}
\label{fig:vgg16-loss}
\end{figure}
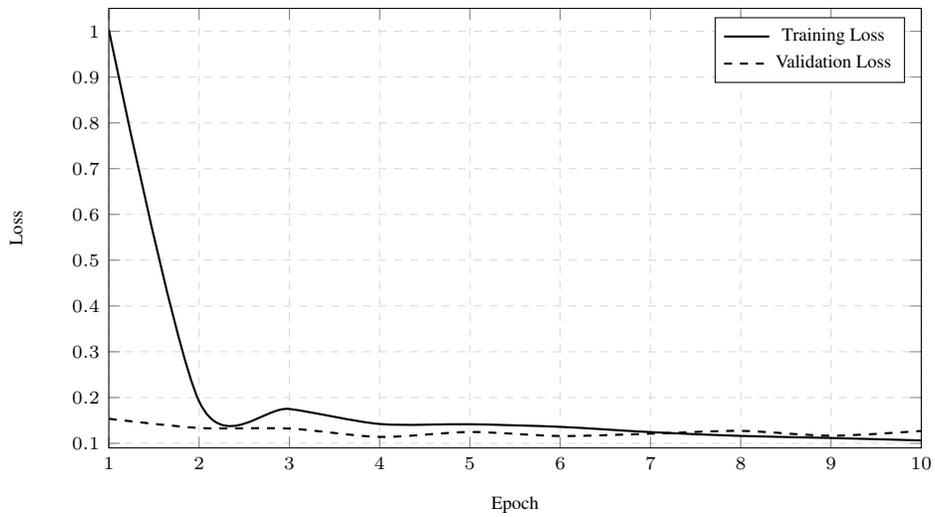

% ===================== FIGURE 3: VGG16 F1 Score =====================
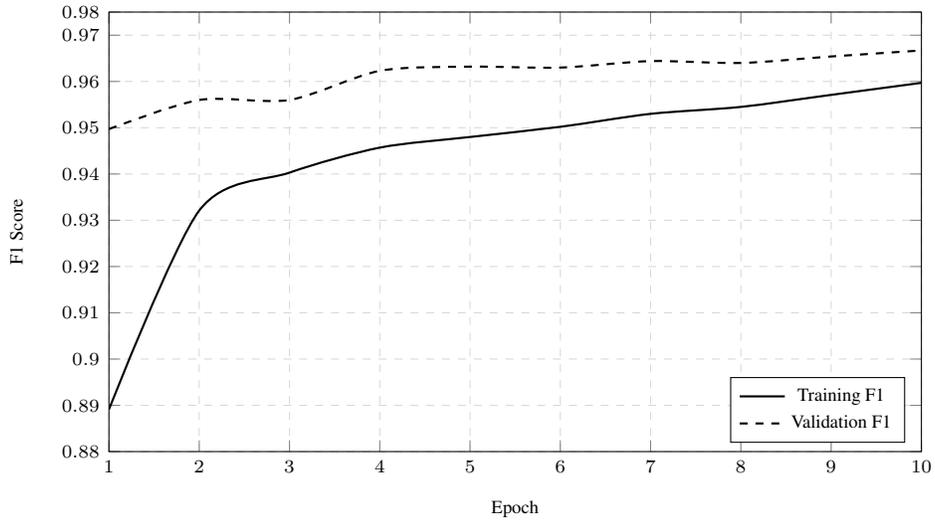
\begin{figure}[htbp]
\centering
\begin{tikzpicture}
\begin{axis}[
    width=0.75\textwidth,
    height=0.45\textwidth,
    xlabel={Epoch},
    ylabel={F1 Score},
    xmin=1, xmax=10,
    ymin=0.88, ymax=0.975,
    xtick={1,2,3,4,5,6,7,8,9,10},
    ytick={0.88,0.89,0.90,0.91,0.92,0.93,0.94,0.95,0.96,0.97,0.975},
    grid=major,
    grid style={dashed,gray!30},
    tick label style={font=\scriptsize},
    label style={font=\scriptsize},
    title style={font=\scriptsize, yshift=1ex},
    legend style={font=\scriptsize, at={(0.98,0.02)}, anchor=south east},
]
\addplot[color=black, thick, solid, smooth, line join=round] coordinates {
    (1,0.8891) (2,0.9321) (3,0.9403) (4,0.9457) (5,0.9480)
    (6,0.9502) (7,0.9530) (8,0.9545) (9,0.9571) (10,0.9597)
};
\addlegendentry{Training F1}

\addplot[color=black, thick, dashed, smooth, line join=round] coordinates {
    (1,0.9497) (2,0.9560) (3,0.9560) (4,0.9623) (5,0.9632)
    (6,0.9630) (7,0.9644) (8,0.9640) (9,0.9654) (10,0.9667)
};
\addlegendentry{Validation F1}
\end{axis}
\end{tikzpicture}
\caption{VGG16 training and validation F1 score across 10 epochs (5-run average).}
\label{fig:vgg16-f1}
\end{figure}

% ===================== FIGURE 4: DenseNet121 Accuracy (5-run average) =====================
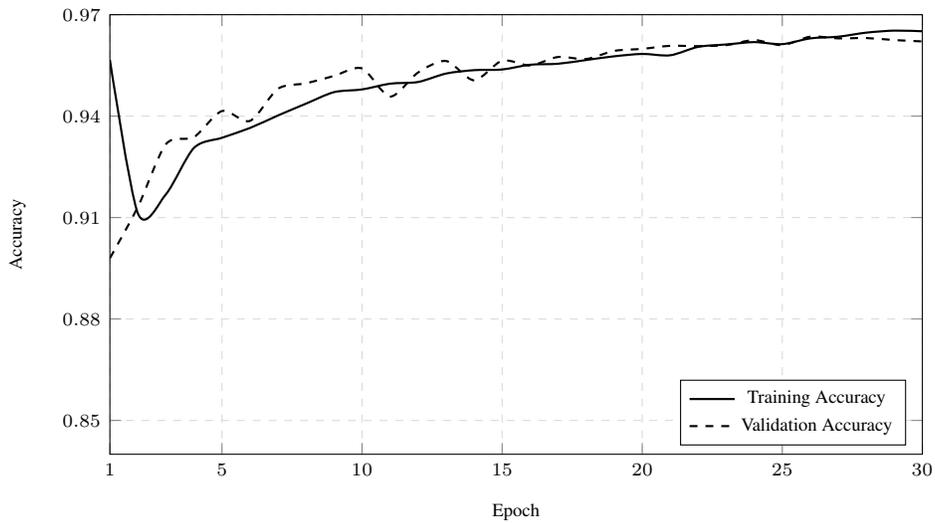
\begin{figure}[htbp]
\centering
\begin{tikzpicture}
\begin{axis}[
    width=0.75\textwidth,
    height=0.45\textwidth,
    xlabel={Epoch},
    ylabel={Accuracy},
    xmin=1, xmax=30,
    ymin=0.84, ymax=0.97,
    xtick={1,5,10,15,20,25,30},
    ytick={0.85,0.88,0.91,0.94,0.97},
    grid=major,
    grid style={dashed,gray!30},
    tick label style={font=\scriptsize},
    label style={font=\scriptsize},
    title style={font=\scriptsize, yshift=1ex},
    legend style={font=\scriptsize, at={(0.98,0.02)}, anchor=south east},
]
\addplot[color=black, thick, solid, smooth, line join=round] coordinates {
(1,0.9567) (2,0.9112) (3,0.9169) (4,0.9306) (5,0.9336)
(6,0.9365) (7,0.9402) (8,0.9437) (9,0.9471) (10,0.9479)
(11,0.9496) (12,0.9501) (13,0.9526) (14,0.9536) (15,0.9538)
(16,0.9552) (17,0.9555) (18,0.9566) (19,0.9577) (20,0.9584)
(21,0.9580) (22,0.9605) (23,0.9612) (24,0.9619) (25,0.9613)
(26,0.9630) (27,0.9635) (28,0.9647) (29,0.9653) (30,0.9651)
};
\addlegendentry{Training Accuracy}

\addplot[color=black, thick, dashed, smooth, line join=round] coordinates {
(1,0.8980) (2,0.9132) (3,0.9317) (4,0.9338) (5,0.9415)
(6,0.9386) (7,0.9481) (8,0.9497) (9,0.9519) (10,0.9541)
(11,0.9458) (12,0.9529) (13,0.9563) (14,0.9506) (15,0.9564)
(16,0.9550) (17,0.9575) (18,0.9569) (19,0.9593) (20,0.9599)
(21,0.9608) (22,0.9607) (23,0.9610) (24,0.9625) (25,0.9610)
(26,0.9635) (27,0.9630) (28,0.9631) (29,0.9625) (30,0.9621)
};
\addlegendentry{Validation Accuracy}
\end{axis}
\end{tikzpicture}
\caption{DenseNet121 training and validation accuracy across 30 epochs (5-run average).}
\label{fig:densenet121-acc}
\end{figure}

% ===================== FIGURE 5: DenseNet121 Loss =====================
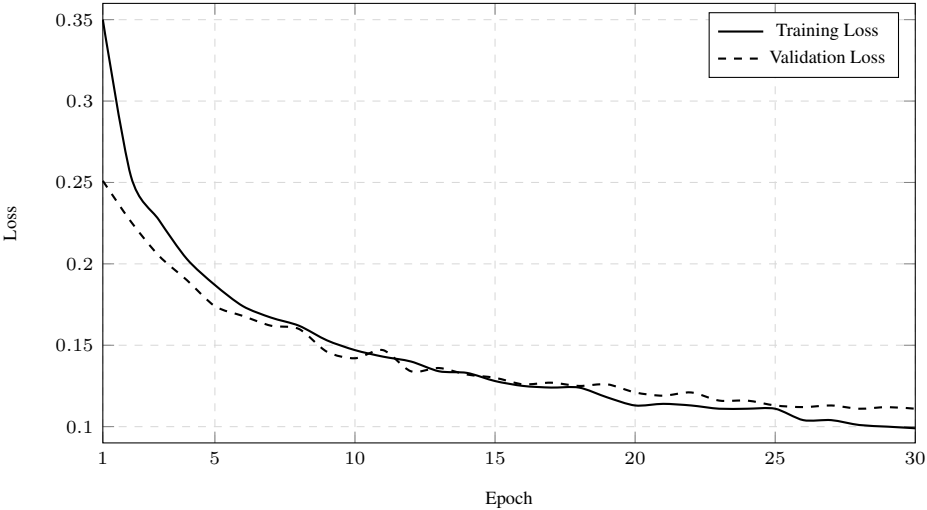
\begin{figure}[htbp]
\centering
\begin{tikzpicture}
\begin{axis}[
    width=0.75\textwidth,
    height=0.45\textwidth,
    xlabel={Epoch},
    ylabel={Loss},
    xmin=1, xmax=30,
    ymin=0.09, ymax=0.36,
    xtick={1,5,10,15,20,25,30},
    ytick={0.1,0.15,0.2,0.25,0.3,0.35},
    grid=major,
    grid style={dashed,gray!30},
    tick label style={font=\scriptsize},
    label style={font=\scriptsize},
    title style={font=\scriptsize, yshift=1ex},
    legend style={font=\scriptsize, at={(0.98,0.98)}, anchor=north east},
]
% Training Loss (5-run average)
\addplot[color=black, thick, solid, smooth, line join=round] coordinates {
(1,0.350) (2,0.254) (3,0.227) (4,0.203) (5,0.187)
(6,0.174) (7,0.167) (8,0.162) (9,0.153) (10,0.147)
(11,0.143) (12,0.140) (13,0.134) (14,0.133) (15,0.128)
(16,0.125) (17,0.124) (18,0.124) (19,0.118) (20,0.113)
(21,0.114) (22,0.113) (23,0.111) (24,0.111) (25,0.111)
(26,0.104) (27,0.104) (28,0.101) (29,0.100) (30,0.099)
};
\addlegendentry{Training Loss}

% Validation Loss (5-run average)
\addplot[color=black, thick, dashed, smooth, line join=round] coordinates {
(1,0.251) (2,0.226) (3,0.205) (4,0.190) (5,0.174)
(6,0.168) (7,0.162) (8,0.160) (9,0.146) (10,0.142)
(11,0.147) (12,0.134) (13,0.136) (14,0.132) (15,0.130)
(16,0.126) (17,0.127) (18,0.125) (19,0.126) (20,0.121)
(21,0.119) (22,0.121) (23,0.116) (24,0.116) (25,0.113)
(26,0.112) (27,0.113) (28,0.111) (29,0.112) (30,0.111)
};
\addlegendentry{Validation Loss}
\end{axis}
\end{tikzpicture}
\caption{DenseNet121 training and validation loss across 30 epochs (5-run average).}
\label{fig:densenet121-loss}
\end{figure}

% ===================== FIGURE 6: DenseNet121 F1 Score =====================
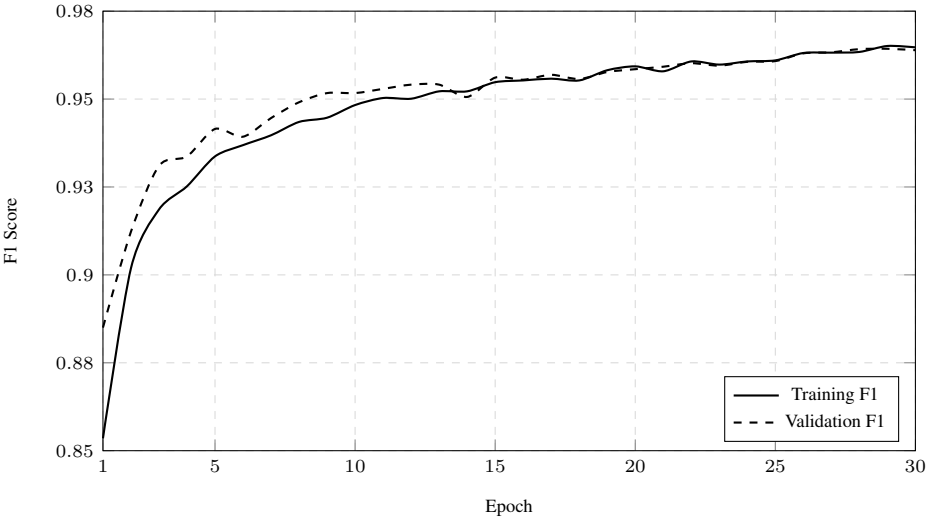
\begin{figure}[htbp]
\centering
\begin{tikzpicture}
\begin{axis}[
    width=0.75\textwidth,
    height=0.45\textwidth,
    xlabel={Epoch},
    ylabel={F1 Score},
    xmin=1, xmax=30,
    ymin=0.85, ymax=0.975,
    xtick={1,5,10,15,20,25,30},
    ytick={0.85,0.875,0.9,0.925,0.95,0.975},
    grid=major,
    grid style={dashed,gray!30},
    tick label style={font=\scriptsize},
    label style={font=\scriptsize},
    title style={font=\scriptsize, yshift=1ex},
    legend style={font=\scriptsize, at={(0.98,0.02)}, anchor=south east},
]
\addplot[color=black, thick, solid, smooth, line join=round] coordinates {
(1,0.8535) (2,0.9019) (3,0.9186) (4,0.9252) (5,0.9337)
(6,0.9369) (7,0.9397) (8,0.9435) (9,0.9447) (10,0.9483)
(11,0.9503) (12,0.9501) (13,0.9522) (14,0.9522) (15,0.9548)
(16,0.9553) (17,0.9558) (18,0.9553) (19,0.9582) (20,0.9593)
(21,0.9579) (22,0.9607) (23,0.9598) (24,0.9607) (25,0.9610)
(26,0.9631) (27,0.9632) (28,0.9634) (29,0.9651) (30,0.9647)
};
\addlegendentry{Training F1}

\addplot[color=black, thick, dashed, smooth, line join=round] coordinates {
(1,0.8850) (2,0.9123) (3,0.9310) (4,0.9337) (5,0.9415)
(6,0.9393) (7,0.9446) (8,0.9491) (9,0.9517) (10,0.9517)
(11,0.9529) (12,0.9541) (13,0.9541) (14,0.9506) (15,0.9561)
(16,0.9555) (17,0.9569) (18,0.9557) (19,0.9577) (20,0.9585)
(21,0.9592) (22,0.9602) (23,0.9595) (24,0.9606) (25,0.9608)
(26,0.9630) (27,0.9633) (28,0.9642) (29,0.9643) (30,0.9639)
};
\addlegendentry{Validation F1}
\end{axis}
\end{tikzpicture}
\caption{DenseNet121 training and validation F1 score across 30 epochs (5-run average).}
\label{fig:densenet121-f1}
\end{figure}

\end{document}